\documentclass[useAMS,usenatbib,usegraphicx]{mn2e}
\usepackage{times}
\newcommand{\reduceme}{\mbox{R\raisebox{-0.35ex}{E}D%
\hspace{-0.05em}\raisebox{0.85ex}{uc}\hspace{-0.90em}%
\raisebox{-.35ex}{{m}}\hspace{0.05em}E}}

\title[MILES: A Medium resolution INT Library of Empirical Spectra]{Medium resolution INT 
Library of Empirical Spectra}
\author[S\'anchez--Bl\'azquez et
  al.]{P. S\'anchez--Bl\'azquez$^{1}$\thanks{E-mail: patricia.sanchezblazquez@epfl.ch}, 
   R. F. Peletier$^{3,4,5}$, 
   J. Jim\'enez-Vicente$^{6}$, 
   N. Cardiel$^{2}$, 
   A. J. Cenarro$^{2}$
\newauthor
   J. Falc\'on--Barroso$^{8,4}$,
   J. Gorgas$^{2}$, 
   S. Selam$^{9}$, 
   A. Vazdekis$^{10}$
\footnotemark[1]\thanks{This file has been amended to
highlight the proper use of \LaTeXe\ code with the class file.}\\
$^{1}$Laboratoire d'Astrophysique, Ecole Polytechnique F\'ed\'erale de Lausanne (EPFL), Observatoire, CH-1290,
Sauverny, Switzerland\\
$^{2}$Departamento de Astrof\'{\i}sica, Universidad Complutense de Madrid, 28040 Madrid, Spain\\
$^{3}$Kapteyn Astronomical Institute, University of Groningen,Postbus 800, 9700 Av Groningen, The Netherlands\\
$^{4}$School of Physics and Astronomy, University of Nottingham, University
Park, Nottingham NG7 2RD, UK\\
$^{5}$Centre de Recherche Astronomique de Lyon, 9 Avenue Charles Andr\'e, 69561 Saint Genis Laval, France\\
$^{6}$Departamento de F\'\i sica Te\'orica y del Cosmos, Universidad de
Granada, Avenida Fuentenueva s/n, 18071 Granada, Spain\\
$^{7}$Calar Alto Observatory, CAHA, Apartado 511, 04044 Almeria, Spain\\
$^{8}$Sterrewacht Leiden, Niels Bohrweg 2, 2333 CA, Leiden, The Netherlands\\
$^{9}$Department of Astronomy and Space Siences, Faculty of Sciences, Ankara University, 06100 Tandogan/Ankara, Turkey\\
$^{10}$Instituto de Astrof\'{\i}sica de Canarias, V\'{\i}a L\'actea s/n, E-38200, La Laguna, Tenerife, Spain}

\begin{document}

\date{Accepted 2006 June13. Received 2006 May 14; in original form
2006 April 6}

\pagerange{\pageref{firstpage}--\pageref{lastpage}} \pubyear{2004}

\maketitle

\label{firstpage}

\begin{abstract}
A new stellar library developed for stellar population synthesis
modeling is presented. The library consist of 985 stars spanning 
a large range in atmospheric parameters. The spectra  were obtained at the 
2.5m INT telescope and  cover the range $\lambda\lambda$
3525--7500~\AA~ at 2.3~\AA~ (FWHM) spectral resolution. 
The spectral resolution, spectral type coverage, flux calibration accuracy and 
number of stars
represent a substantial improvement over previous libraries used in
population synthesis models.
\end{abstract}


\section{Introduction}
  With this paper we start a project aimed at improving the existing
  tools for extracting stellar population information using the
  optical region of composite spectra.  Although the main motivation
  of this work is to use this new calibration to study the stellar
  content of galaxies using spectra of unresolved stellar populations, we expect that the material
  presented here and in future papers will be useful in other areas of
  astronomy. This includes a new stellar library, a set of homogenous
  atmospheric parameters, a re-definition and re-calibration of
  spectral line indices, empirical fitting functions describing the
  behavior of indices with stellar parameters, and stellar population
  model predictions.
   
 A comprehensive spectral library with medium-to-high resolution and a
 good coverage of the Hertzsprung-Russell (HR) diagram is an essential
 tool in several areas of astronomy. In particular, this is one of the
 most important ingredients of stellar population synthesis, providing
 the behavior of individual stellar spectra as a function of temperature, 
 gravity
 and chemical abundances.  Unfortunately, the empirical libraries
 included in this kind of models up to now contained few stars
 with non-solar metallicities, compromising the accuracy of
 predictions at low and high metallicities.

 This problem has usually been partially solved by using empirical
 fitting functions, polynomials that relate the stellar atmospheric parameters
 ($T_{\rm eff}$, $\log g$, and [Fe/H]) to measured equivalent
 widths (e.g. Gorgas et al. 1993; Worthey et al. 1994, Worthey \& Ottaviani 1997). These
 functions allow the inclusion of any star required by the model (but within
 the stellar atmospheric parameter ranges covered by the functions) using a smooth
 interpolation.  However, the new generation of stellar population
 models goes beyond the prediction of individual features for a simple
 stellar population (SSP), and they attempt to synthesize full
 spectral energy distributions (SED) (Vazdekis 1999; Vazdekis et
 al. 2003; Bruzual \& Charlot 2003). In this case, the fitting
 functions can not be used, and a library of stars covering the 
 full range of atmospheric
 parameters in an ample and homogeneous way is urgently demanded.
 Moreover, although the evolutionary synthesis codes do not require
 absolute fluxes,the different stellar spectra must
 be properly flux calibrated in a relative sense so that the whole spectral 
 energy distribution can be modeled. This, however, is
 quite difficult to
 achieve in practice, due to the wavelength dependent flux losses caused by
 differential refraction when a narrow slit is used in order to obtain
 a fair spectral resolution.

Another important caveat in the interpretation of the composite
spectrum of a given galaxy is the difficulty of disentangling the
effects of age and metallicity (eg. Worthey 1994). Due to blending
effects, this problem is worsened when working at low spectral
resolution, as it is the case when low resolution stellar libraries
are used (eg. Worthey et al. 1994; Gunn \& Stryker 1983).  There are
a few studies that have attempted to include spectra features at
higher resolution (Rose 1994; Jones \& Worthey 1995; Vazdekis \&
Arimoto 1999). However, predicting such high-dispersion SEDs is very
difficult owing to the unavailability of a library with 
the required input spectra.

Whilst the new generation of large telescopes is already gathering
high quality spectra for low and high redshift galaxies, the stellar
population models suffer from a lack of extensive empirical stellar
libraries to successfully interpret the observational data. At the
moment, the available stellar libraries have important shortcomings
such as small number of stars, poor coverage of atmospheric
parameters, narrow spectral ranges, low resolution and non flux-calibrated response
curves. Here we present
a library that overcomes some of the limitations of the previous
ones. The new library, at spectral resolution of 2.3~\AA~(FWHM),
contains 985 stars with metallicities ranging from  [Fe/H] $\sim
-2.7$ to $+1.0$ and a wide range of temperatures.

 The outline of the paper is the following. In order to justify the
 observation of a new stellar library, in Section 2 we review previous
 libraries in the optical spectral region. Section 3 presents the
 criteria to select the sample and, in Section 4, the observations and
 data reduction are summarized. Section 5 presents the library, while
 a quality control and comparison with spectra from previous libraries
 are given in Sections 6 and 7 respectively.  Finally, Section 8
 summarizes the main results of this paper.

 \section{Previous stellar libraries in the optical region} 

Table \ref{otras-librerias} shows some of the previous libraries in
the blue spectral range with their principal characteristics. We only
include those libraries which have been used or have been built for
stellar population synthesis purposes. In the following paragraphs we
comment on the main advantages and caveats of a selected subsample.
 
 The most widely used library up to now has been the Lick/IDS library
 (Gorgas et al. 1993; Worthey et al. 1994), which contains about 430
 stars in the spectral range $\lambda\lambda$4000--6200~\AA~. Examples
 of population models using this library are those of Bruzual \&
 Charlot (1993); Worthey (1994); Vazdekis et al. (1996); Thomas, Maraston \&
 Bender (2003); Thomas, Maraston \& Korn (2004). The Lick
 library, based on observations taken in the 1970's and 1980's with the IDS, a
 photon counting device, has been very useful, since it contains stars with a fair
 range of $T_{\rm eff}$, $\log g$, and [Fe/H]. However, a number of
 well-known problems is inherent to this library.  Since the stars are
 not properly flux calibrated, the use of the predictions on the Lick
 system requires a proper conversion of the observational data to the
 instrumental response curve of the original dataset (see the analysis
 by Worthey \& Ottaviani 1997).  This is usually done by observing a
 number of Lick stars with the same instrumental configuration as the
 one used for the galaxy. Then, by comparing with the tabulated Lick
 measurements, one can find empirical correction factors for each
 individual absorption feature. Another important step to be followed
 is the pre-broadening of the observational spectra to match the
 resolution of Lick/IDS, which suffers from an ill-defined
 wavelength dependence. Note that this means that part of the
 information contained in high resolution galaxy spectra is lost.
 Furthermore, the spectra of this library have a low effective signal-to-noise
 ratio due to the significant flat-field-noise (Dalle Ore et al. 1991; Worthey 
 et al. 1994; Trager et al. 1998). This translates into very large
 systematic errors in the indices much larger than present-day 
 galaxy data. In fact, the accuracy of the measurements based in the Lick system 
 is often limited by the stellar library, rather
 than by the galaxy data.

With the
availability of new and improved stellar libraries, a new generation of
stellar population models are able to reproduce galaxy spectra and not
just line strengths.
Jones' library (Jones
 1997) was the first to provide flux calibrated spectra with a
 moderately high spectral resolution (1.8~\AA).  Using this library,
 Vazdekis (1999) presented, for the first time, stellar population
 synthesis models predicting the whole spectrum of a single stellar
 population. Another example of a population synthesis model using
 this library is Schiavon et al. (2002).  However, Jones' library is
 limited to two narrow wavelength regions, 3820--4500 and 4780--5460~
 \AA, and it is sparse in dwarfs hotter than $\sim$ 7000 K and
 metal-poor giants ([Fe/H]~$\leq -0.5$). 
 The first limitation prevents stellar population models from predicting
populations younger than 4 Gyr, while the second limitation affects the 
models of old, metal poor systems like globular clusters.
 More recently,
 a new stellar library at very high spectral resolution (0.1~\AA), and
 covering a much larger wavelength range (4100--6800~\AA), has become
 available (ELODIE; Prugniel \& Soubiran 2001). The physical parameter range
 of this library is limited, and the flux calibration is compromised by
 the use of an echelle spectrograph.
 Recently, an updated version of ELODIE stellar library (Prugniel \& Soubiran
 2004, ELODIE.3 hereafter)
 has been also incorporated into a new population synthesis models (Le Borgne
 et al. 2004). This version doubles the size of the previous one and 
 offer an improved coverage of atmospheric parameters. 

 Bruzual \& Charlot (2003) have recently presented new stellar
 populations synthesis models at the resolution of 3~\AA\ (FWHM) across
 the wavelength range from 3200 to 9500~\AA. Their predictions are
 based on a new library (STELIB) of observed stellar spectra recently
 assembled by Le Borgne et al. (2003). This library represents a
 substantial improvement over previous libraries commonly used in
 population synthesis models. However, the sample needs some
 completion for extreme metallicities and the number of stars is not
 very high. In particular, the main problem of this library is the lack
 of metal-rich giants stars.

 Finally, near completion of the present paper a new stellar library  
 (INDO-US; Valdes et al. 2004) was published. 
 This library
 contains a large number of stars (1273) covering a fair range in
 atmospheric parameters. Unfortunately, the authors could not obtain
 accurate spectrophotometry but they fit each observation to a
 spectral energy distribution standard with a close match in spectral
 type using the compilation of Pickles (1998).

 To summarize, the quality of available stellar libraries for
 population synthesis has improved remarkably over the last
 years. However, a single library with simultaneous fair spectral
 resolution (eg. ELODIE), atmospheric parameter coverage
 (eg. INDO-US), wide spectral range (eg. STELIB) and with an accurate
 flux calibration  is still lacking.  In the following
 sections we will compare the new library (MILES) with the
 previous ones in some of the above relevant characteristics.
 
 \begin{table*}
 \centering
  \caption{Some of the previous libraries in the optical region devoted to stellar population studies. \label{otras-librerias}} 
 \begin{tabular}{l l l l l}
 \hline
 Reference & Resolution& Spectral range & Number of stars & Comments\\
           & FWHM(\AA)  &               &                 &         \\
 \hline
 Spinrad (1962)                    &                 &                 &     & Spectrophotometry\\
 Spinrad \& Taylor (1971)            &                &                  &    & Spectrophotometry \\
 Gunn \& Stryker (1983)            &20-40     & 3130--10800~\AA  & 175 & \\     
 Kitt Peak (Jacoby et al. 1984)     & 4.5      & 3510--7427~\AA  & 161  & Only solar metallicity\\
 Pickles (1985)                    &10-17     & 3600--1000~\AA  & 200 & Solar metallicity except G-K giants  \\
 Lick/IDS (Worthey et al. 1994)    &9-11      & 4100--6300~\AA  & 425 & Not
 flux calibrated, variable resolution  \\
 Kirkpatrick et al. (1991)         & 8/18     & 6300--9000~\AA  & 39   & No atmospheric correction \\
 Silva \& Cornell (1992)           & 11       & 3510--8930~\AA  & 72 groups& Poor metallicity coverage\\
 Serote Roos et al. (1996)         &1.25      & 4800--9000~\AA  & 21   &   \\          
 Jones (1997)                      &1.8       & 3856--4476~\AA  & 684 &Flux calibrated \\
                                   &          & 4795--5465~\AA  &     &                \\
 Pickles (1998)                    &          & 1150--10620~\AA & 131 groups  & Flux calibrated\\
 ELODIE (Prugniel \& Soubiran 2001)&0.1       & 4100--6800~\AA  &709 & Echelle   \\
 STELIB (Le Borgne et al. 2003)    &3.0       & 3200--9500~\AA  & 249 & Flux calibrated\\
 INDO-US (Valdes et al. 2004)      &1.0       & 3460--9464~\AA  & 1273& Poor flux calibrated \\ 
\hline
MILES &2.3 & 3525--7500~\AA & 995 & \\
\hline

\end{tabular}
 \end{table*}

 \section{Sample selection}

\begin{figure}
   \resizebox{1.0\hsize}{!}{\includegraphics[angle=-90]{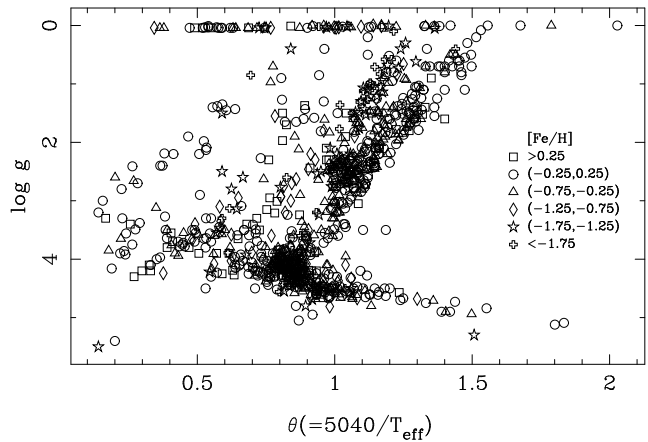}} 
 \caption{Gravity-temperature diagram for the library stars. Different symbols are
 used to indicate stars of different metallicities, as shown in the key. \label{fig-hrdiagram}} 
 \end{figure}

Although the new library is expected to have different applications,
the selection of the stars is optimized for their inclusion in stellar
population models.  Figure \ref{fig-hrdiagram} shows a pseudo-HR
diagram for the whole sample. MILES includes 232 of the 424 stars with
known atmospheric parameters of the Lick/IDS library (Burstein et
al. 1984; Faber et al. 1985; Burstein, Faber \& Gonz\'alez 1986;
Gorgas et al. 1993; Worthey et al. 1994). The atmospheric parameter
coverage of this subsample is representative of that library, and
spans a wide range in spectral types and luminosity classes. Most of
them are field stars from the solar neighborhood, but stars covering a
wide range in age (from open clusters) and with different
metallicities (from galactic globular clusters) are also included.  In
addition, with the aim of filling gaps and enlarging the parameter
space coverage, stars from additional compilations were carefully
selected (see below).

G8--K0 metal rich stars ($+0.02<{\rm [Fe/H]}<+0.5$) with temperatures
between 5200 and 5500 K were extracted from Castro et al. (1997),
Feltzing \& Gustafsson (1998), Randich et al. (1999), Sadakane et
al. (1999), and Thor\'en \& Feltzing (2000). We also obtained some stars
from a list kindly provided by K. Fuhrmann (private communication). We
also added to the sample some stars with temperatures above 6000 K and
metallicities higher than $+0.2$ (from Gonz\'alez \& Laws 2000), which
will allow to reduce the uncertainties in the predictions of our models
at this metallicity.

The inclusion of hot dwarf stars with low metallicities are essential
to predict the turn-off of the main sequence due to their high
contribution to the total light. We obtained these stars from Cayrel
de Strobel et al. (1997).
   
MILES also contains dwarf stars with temperatures below 5000 K. These
stars, which were absent in the Lick library, allow to make
predictions using IMF with high slopes, and have been obtained from
Kollatschny (1980), McWilliam (1990), Castro et al. (1997), Favata,
Micela \& Sciortino (1997), Mallik (1998), Perrin et al. (1998),
Zboril \& Byrne (1998), Randich et al. (1999), and Thor\'en \& Feltzing
(2000).
 
We also included 17 stars to the region of the diagram corresponding
to cool and metal rich (with [Fe/H]~$>+0.15$) giants stars from
McWilliam (1990), Ram\'{\i}rez et al. (2000), and
Fern\'andez--Villaca\~nas et al. (1990). Some metal poor giant stars
with $T_{\rm eff}<6000$~K, from McWilliam (1990), were also
incorporated in order to improve the predictions of old stellar
populations and to study the effect of the horizontal branch.

In the selection of the sample we have tried to minimize the inclusion
of spectroscopic binaries, peculiar stars, stars with chromospheric
emission and stars with strong variability in regions of the HR diagram
where stars are not expected to vary significantly.
 For this purpose, we used SIMBAD and the
Kholopov et al. (1998) database of variable stars.
 
\begin{figure*}
  \resizebox{1.0\hsize}{!}{\includegraphics[angle=-90]{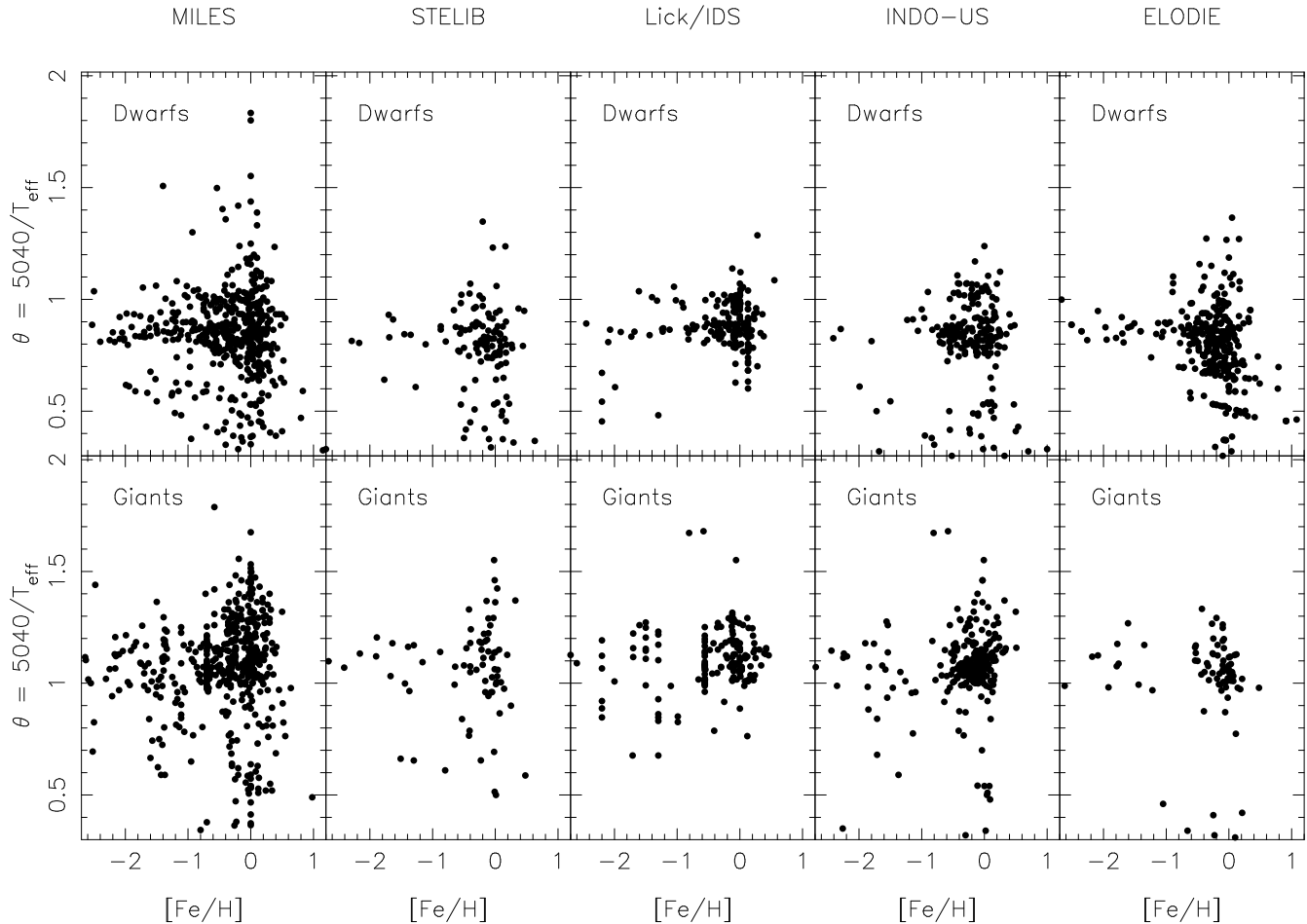}}
 \caption{Parameter coverage of MILES (left panel) compared with other stellar libraries.
 \label{fig-comparison}}
 \end{figure*}

Fig. \ref{fig-comparison} shows the atmospheric parameters coverage of
MILES compared with other libraries. As can be seen, the numbers of
cool and super metal rich stars, metal poor stars, and hot stars
($T_{\rm eff}> 6500$~K) have been greatly enhanced with respect to
previous works.


 \section{Observations and data reduction}

 The spectra of the stellar library were obtained during a total 25
 nights in five observing runs from 2000 to 2001 using the 2.5~m INT at the
 Roque de los Muchachos Observatory (La Palma, Spain).  All the stars
 were observed with the same instrumental configuration, which ensures
 a high homogeneity among the data.

  Each star was observed with three different setups, two of them
  devoted to obtain the red and the blue part of the spectra, and a
  third one, with a wide slit (6 arcsec) and a low dispersion grating
  (WIDE hereafter), which was acquired to ensure a fair flux
  calibration avoiding selective flux losses due to the  atmospheric
  differential refraction.  A description of these and other
  instrumental details is given in Table \ref{intru-configuration}.
  Typical exposure times varied from a few seconds for bright stars to
  1800~s for the faintest cluster stars. These provided typical values
  of SN(\AA) (signal-to-noise ratio per angstrom) averaged over the whole 
  spectral range of $\sim150$ 
  for field and open cluster stars, and $\sim 50$ for globular
  cluster stars.
  
 \begin{table}
 \centering
 \begin{tabular}{l l l r  }
 \hline
      & Grating & Detector & Dispersion \\
   \hline
   \hline
     Red  &  R300V & EEV10& 0.9~\AA/pixel  \\
     Blue &  R150V & EEV10 &0.9~\AA/pixel  \\
     Wide &  R632V & EEV10 & 1.86~\AA/pixel \\
     \hline
          & Slit width& Spectral coverage&Filter\\
     \hline
      \hline
     Red  &   0.7$''$& 3500-5630&GG495\\
     Blue &   0.7$''$ & 5000-7500&none\\
     Wide &   6.0$''$& 3350-7500&WG360\\
 \hline
 \end{tabular}
 \caption{Observational configurations. ``Wide'' refers to the configuration with the wide slit.
 \label{intru-configuration}}
 \end{table}

 The basic data reduction was performed with IRAF\footnote{IRAF is
 distributed by the National Optical Astronomy Observatories, USA,
 which are operated by the Association of Universities for Research in
 Astronomy, Inc., under cooperative agreement with the National
 Science Foundation, USA.} and
 \reduceme\footnote{http://www.ucm.es/info/Astrof/software/reduceme/reduceme.html}
 (Cardiel 1999). \reduceme~ allows a parallel treatment of data and
 error frames and, therefore, produces an associated error spectrum
 for each individual data spectrum. We carried out a standard
 reduction procedure for spectroscopic data: bias and dark
 subtraction, cosmic ray cleaning, flat-fielding, C-distortion  (geometrical 
 distortion of the image along the spatial direction)
 correction, wavelength calibration, S-distortion  (geometrical distortion 
 of the image along the spectral direction) correction, sky
 subtraction, spectrum extraction and relative flux calibration.
 Atmospheric extinction correction was applied to all the spectra using 
 wavelength dependent extinction curves provided by the observatory
  (King 1985, http://www.ing.iac.es).
 Some of the reduction steps that required more careful work are explained in
 detail in the following subsections.
 \subsection{Wavelength calibration}

 Arc spectra from Cu-Ar, Cu-Ne and Cu-N lamps were acquired to perform
 the wavelength calibration. The typical number of lines used ranged
 from 70 to 100. In order to optimize the observing time, we did
 not acquire comparison arc frames for each individual exposure of a
 library star but only for a previously selected subsample of stars
 covering all the spectral types and luminosity classes in each
 run. The selected spectra were wavelength calibrated with their own
 arc exposures taking into account their radial velocities, whereas
 the calibration of any other star was performed by a comparison with
 the most similar, already calibrated, reference spectrum. This
 working procedure is based on the expected constancy of the
 functional form of the wavelength calibration polynomial within a
 considered observing run. In this sense, the algorithm that we used
 is as follows: after applying a test $x-$shift (in pixels) to any
 previous wavelength calibration polynomial, we obtained a new
 polynomial which was used to calibrate the spectrum. Next, the
 calibrated spectrum was corrected from its own radial velocity and,
 finally, the spectrum was cross-correlated with a reference spectrum
 of similar spectral type and luminosity class, in order to derive the
 wavelength offset between both spectra. By repeating this procedure,
 it is possible to obtain the dependence of the wavelength offset as a
 function of the test $x-$shift and, as a consequence, to derive the
 required $x-$shift corresponding to a null wavelength offset. The RMS
 dispersion of the residuals is of the order of 0.1~\AA.

  Once the wavelength calibration procedure was applied to the whole
  star sample, we still found small shifts due to uncertainties in the
  published radial velocities. In order to correct for this effect,
  each star was cross-correlated with the high resolution solar
  spectrum obtained from BASS2000 (BAse de donnees
  Solaire Sol; http://bass2000.obspm.fr) in the wavelength region of the 
  Ca H and K lines. First the solar spectrum was cross-correlated 
  with the stars that were cool enough to have the H and K lines.
  The derived shifts were then applied to the corresponding spectra. 
  As a next step we then cross-correlated the hotter stars with stars that 
  did exhibit simultaneously Ca H and K, and Balmer lines, and applied the shifts. 
 
\subsection{Spectrum extraction}
 A first extraction of the spectra was performed 
 adding the number of scans which maximised the S/N. 
 However, in these spectra,    
 the presence of scattered light was weakening the 
 spectral lines. Scattered light in the spectrograph
 results from undesired reflections from the refractive optics
 and the CCD, imperfections in the reflective surfaces and scattering 
 of the light outside first order from the spectrograph case.
 Scattered light amounting to 6\% of the dispersed light is scattered
 fairly uniformly across the CCD surface, affecting more strongly 
 to the spectra with low level signal. 
  To minimize  the uncertainties due to scatterd light we extracted the spectra adding
  only  3 scans (the central and one more to each side).
 Although, in principle, the extraction
 of a reduce number of spectra may affect the shape of the 
 continuum (due to the differential refraction in the atmosphere), 
 we performed the flux calibration using a different set of 
 stars observed with a slit of 6 arcsec (see next section). Therefore, 
 the accuracy of the flux calibration is not compromised by 
 this extraction. 

\subsection{Flux calibration. The second order problem}
 One of the major problems of the Lick/IDS library for computing
 spectra from stellar population models is that the stars
 are not properly flux calibrated. Therefore, the use of model predictions based
 on that system requires a proper conversion of the
 observational data to the characteristics of the instrumental IDS
 response curve (see discussion by Worthey \& Ottaviani
 1997). Also, a properly flux calibrated stellar library is essential
 to derive reliable predictions for the whole spectrum, and not only for
 individual features (Vazdekis 1999; Bruzual \& Charlot 2003). It must
 be noted that we have not attempted to obtain absolute fluxes since
 both, the evolutionary synthesis code and the line-strength indices,
 only require relative fluxes. 

In order to perform a reliable flux calibration several
spectrophotometric standards (BD+33~2642, \mbox{G 60-54}, BD+28~4211,
HD~93521 and BD+75~325) were observed along each night at different
air-masses.  A special effort was made to avoid the selective flux
losses due to the differential refraction. For this reason, all
stars were also observed through a 6$''$ slit. This additional
spectrum was flux calibrated using the standard procedure and the derived
continuum shape was then imposed on the two high resolution spectra.

 In spite of using a colour filter, the red end ($\lambda >$ 6700~\AA)
 of the low resolution spectra suffered from second order
 contamination. Fortunately, since the low resolution spectra begin at
 3350~\AA, it was possible to correct from this contamination. To do
 that, we made use of two different standard stars, S$_{a}$ and
 S$_{b}$. The observed spectra of the standard stars before flux
 calibrating can be expressed as:
 \begin{equation}
 S_{a}=C_{1}T_{a}+C_{2}T_{2a}\\
 S_{b}=C_{1}T_{b}+C_{2}T_{2b}
 \label{equtont}
 \end{equation}
 where $T_{a}$ and $T_{b}$ are the tabulated data for $S_{a}$ and
 $S_{b}$ respectively and $T_{2a}$ and $T_{2b}$ are these
 re-sampled to twice the resolution and displaced 
 3350~\AA. $C_{1}$ and $C_{2}$ represent the instrumental response for
 the light issuing from the first and second dispersion orders,
 respectively. Solving the system of equations (\ref{equtont}) we obtain the response curves
 as:\\ 
\begin{equation}
 C_{2}=\frac{T_{a}S_{b}-T_{b}S_{a}}{T_{2b}T_{a}-T_{2a}T_{b}}
 \end{equation} and \begin{equation}
 C_{1}=\frac{S_{a}-C_{2}T_{2a}}{T_{a}} \end{equation}

 With these response curves, we obtain the flux calibrated star from 3350 to 6700~\AA~ as:\\
 \begin{equation}
 S_{a}'(\lambda<6700\rm{\AA})=\frac{S_{a}}{C_{1}}
 \end{equation}
and after, resampling $S_{a}'$ to a double dispersion and
shifting the spectra by 6700~\AA~ (we refer to the resulting spectra of
these operations as $S_{2a}'$), we finally obtain the whole
calibrated spectrum as:\\
 \begin{equation}
 S_{a}'=\frac{S_{a}-C_{2}S'_{2a}}{C_{1}}
 \end{equation}

\begin{figure}
 \resizebox{1.0\hsize}{!}{\includegraphics[angle=-90]{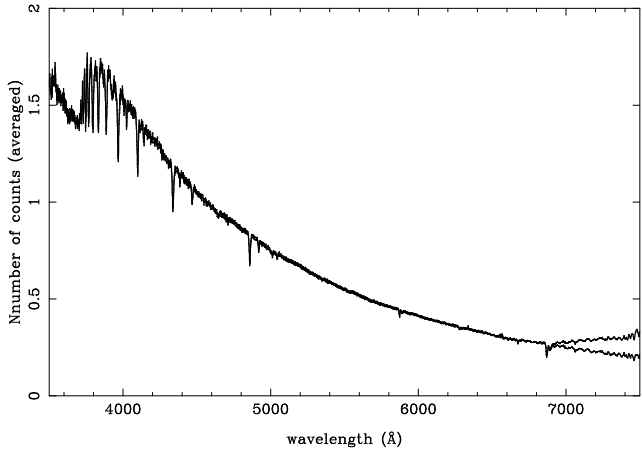}}
 \caption{Standard star spectrum after and before correcting for second order contamination.\label{fig-secondorder}}
 \end{figure}

 Figure \ref{fig-secondorder} shows the flux calibrated spectrum of
 the standard star BD+33~2642 before and after correcting for this
 second order contamination with the above procedure.
 
\subsection{The spectral resolution}

\begin{figure}
 \resizebox{1.0\hsize}{!}{\includegraphics{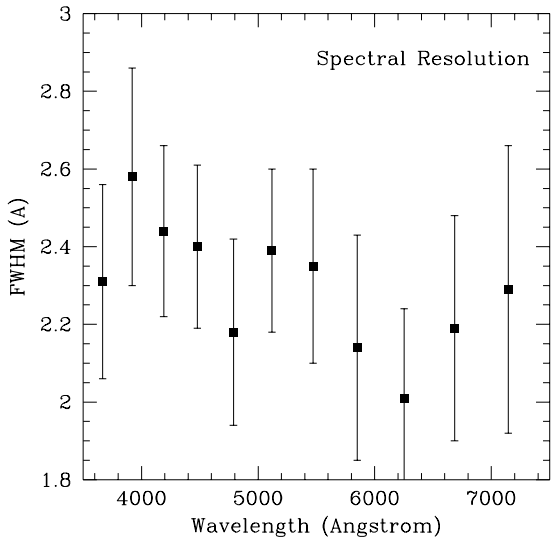}}
 \caption{Mean FWHM (expressed in  \AA) of MILES spectra measured in 
 11 different spectral regions.
The error bars indicate the RMS dispersion of the values measured 
with all the stars of the library.
\label{focus-variacion}} 
\end{figure}

 Since it is important to know the spectral resolution of MILES, we first
 used the different calibration lamp spectra to homogenise the spectral
 resolution of the stars. After that, we selected a set of 6 stars from the 
 INDO-US library (Valdes et al.\ 2004), and fitted a linear combination
 of these stars to the spectrum of every star in 11 different wavelength 
 regions. During the process, the resolution of the MILES-spectra was 
 obtained by determining the best-fitting Gaussian with which the linear
 combination of INDO-US spectra had to be convolved, and correcting for 
 the intrinsic width of the INDO-US spectra (checked to be 1.0\AA for
 HD 38007, a G0V star, similar to the Sun, by comparing with a high resolution 
 solar spectrum). To do that, we used the task ppxf (Cappellari \& Emsellem
 2004). The average resolution of the stars is given in
 Fig.~\ref{focus-variacion}. The figure shows that the resolution amounts
 to 2.3$\pm$0.1 \AA.
 \section{The final spectra and the database}
 \label{sec.finalspectra}

\begin{figure}
 \resizebox{1.0\hsize}{!}{\includegraphics[angle=-90]{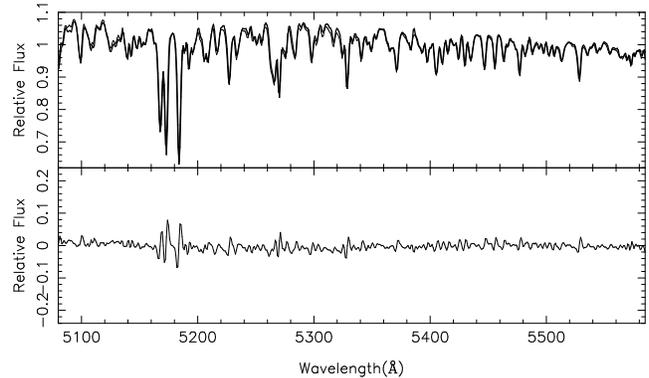}}
 \caption{RED (thick line) and BLUE (thin line) spectra of the star HD157214 in 
the spectral range in common between the two instrumental configurations.
\label{fig-combinedspectra}}
 \end{figure}

At the end of the reduction, spectra in the red and blue spectral ranges
were combined to produce a unique high-resolution flux-calibrated
spectrum for each star, together with a corresponding error spectrum.
The spectra in the two wavelength ranges share a common spectral range (from 5000 to 5630 
~\AA) in which a mean spectrum was computed by performing an error-weighted average.
In Figure \ref{fig-combinedspectra} we show a
typical example of the match between the red and the blue spectra
in this wavelength interval.  If we take into
account that the two spectra have been calibrated independently,
the agreement is very good. This gives support to the quality of the
reduction process.

Finally, all the stellar spectra were corrected for interstellar
reddening using Fitzpatrick (1999) reddening law. For 444 stars,
$E(B-V)$ values were taken from Savage et al. (1985), Friedemann (1992),
Silva \& Cornell (1992), Gorgas et al. (1993), Carney et al. (1994),
Snow et al. (1994), Alonso, Arribas, \& Mart\'{\i}nez--Roger (1996),
Dyck et al. (1996), Harris (1996), Schuster et al. (1996), Twarog et
al. (1997), Taylor (1999), Beers et al. (1999), Dias et al. (2002),
Stetson, Bruntt, \& Grundahl (2003), and V.  Vansevi\^cius (private
communication).

For stars lacking $E(B-V)$ estimates, we obtained new values following
the procedure described in Appendix~A. With this method we achieved
reddenings for 275 new stars. For the remaining 284 stars without
reddening determinations these were estimated as follows:

For 51 stars, $E(B-V)$ values were calculated following Schuster et
al. (1996) from $uvby$-$\beta$ photometry obtained from Hauck \&
Mermilliod (1998). The RMS dispersion between the values obtained with this
procedure and those published in the literature (see previous
references) is 0.012 mag. For another subsample of 41 stars, reddenings
were calculated with the calibration by Bonifacio, Caffau, \& Molaro
(2000) using synthetic broad-band Johnson colours and the line indices KP
and HP2 measured directly over our spectra. The agreement between
$E(B-V)$ values obtained with this procedure and the literature values
is within 0.05 magnitudes.  $E(B-V)$ values for 72 stars were
calculated following Janes (1997) using DDO photometry; C(45-48) and
C(42-45) were also measured in our spectra. The RMS dispersion between the
values obtained with this method and those from the literature is 0.029
magnitudes.  Despite all these efforts, 145 stars still lacked reddening
determinations. For these stars, $E(B-V)$ values were calculated from
their Galactic coordinates and parallaxes by adopting the extinction
model by Chen et al. (1999), based on the COBE/IRAS all sky reddening
map (Schlegel, Finkbeiner, \& Davies 1998). The agreement between the 
 $E(B-V)$ values derived in this way and the values extracted from the literature
 is within 0.048 magnitudes.

 Table \ref{excesos.metodos} summarizes the different methods of 
 extinction determinations in order of preference.
 The last column of the table 
 contains the number of $E(B-V)$ obtained with each method.

 \begin{table}
 \centering
 \begin{tabular}{l rr}
 Method                  & $\sigma$  & Number\\
 \hline
 1) Literature              &           &  444 \\
 2) Appendix                & 0.032     &  275 \\
 3) Jones (1997)            & 0.029     &  72  \\
 4) Schuster et al. (1996)  & 0.012     &  51  \\
 5) Bonifacio et al. (2000) & 0.050     &  41  \\
 6) Extinction Maps         & 0.048     &  145\\
 \hline
 \end{tabular}
\caption{Methods applied to obtain reddenings for the 
 library stars, listed in order
 of preference. Second column shows the RMS dispersion in the 
 comparison of each method with the values collected from the 
 literature. Third column shows the number of stars with 
 $E(B-V)$ determined with each different procedure.
 \label{excesos.metodos}}
 \end{table}

\begin{figure*}
\resizebox{1.0\hsize}{!}{\includegraphics[angle=-90]{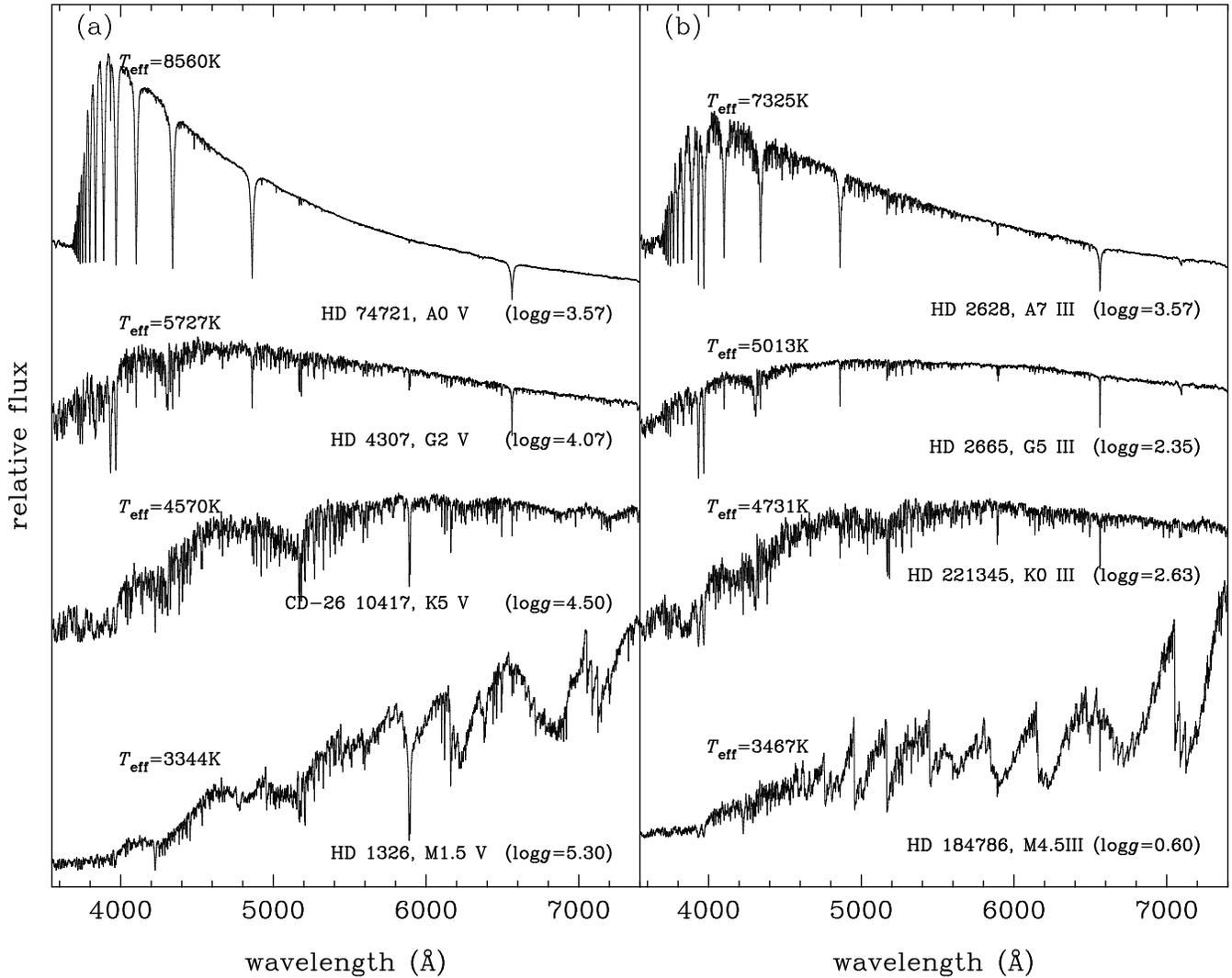}}
\caption{Sequences of spectral types for a sample of (a) dwarf and (b) giant stars from the library. 
Effective temperatures, names, spectral types and surface gravities are given in the labels.
\label{fig-espectra}}
 \end{figure*} 

As an example of final spectra, Figure~\ref{fig-espectra} shows
comparative sequences of spectral types for a sample of dwarf and
giant stars from the library. Information for each star in the
database is presented in Table \ref{tabla.final}, available in the
electronic edition and at the Library web site ( http://www.ucm.es/info/Astrof/
MILES/miles.html). Individual spectra for the complete library are
available at the same www page.  The atmospheric parameters have been
derived from values in the literature, transforming them to a
homogeneous system following Cenarro et al. (2001). The detailed
description of this method will be given in a forthcoming paper (Peletier
et al. 2006; in preparation). The last two columns of Table
\ref{tabla.final} show the adopted $E(B-V)$ values and the reference
from which they were obtained (see the electronic version of the table
for an explanation of the reference codes). When several references
are marked, an averaged value has been adopted.

 \begin{table*}
 \begin{tabular}{l l l l l l l l l}
 Star         & RA (J2000.0)  & DEC (J2000.0)  & Sp. Type &  $T_{\rm eff}$ & $\log g$ & 
[Fe/H] & $E(B-V)$ &Ref\\
\hline
\hline
 BD +00 2058  &07:43:43.96&-00:04:00.9& sd:F   &  6024          &4.50     &$-1.56$   & 0.020  & 4,5     \\
 BD +01 2916  &14:21:45.26&+00:46:59.2&  K0    &  4238          &0.34     &$-1.49$   & 0.030  & 6      \\
 BD +04 4551  &20:48:50.72&+05:11:58.8&  F7Vw  &  5770          &3.87     &$-1.62$   & 0.000  & 8     \\
 BD +05 3080  &15:45:52.40&+05:02:26.6&  K2    &  5016          &4.00     &$-0.79$   & 0.000  & 4,6     \\
 BD +06 0648  &04:13:13.11&+06:36:01.7&  K0    &  4400          &1.03     &$-2.12$   & 0.000  & 1,7      \\
 BD +06 2986  &15:04:53.53&+05:38:17.1&  K5    &  4450          &4.80     &$-0.30$   & 0.006  & 24      \\
 BD +09 0352  &02:41:13.64&+09:46:12.1&  F2    &  5894          &4.44     &$-2.20$   & 0.020  & 4       \\
 BD +09 2190  &09:29:15.56&+08:38:00.5&  A0    &  6316          &4.56     &$-2.71$   & 0.010  & 4,6     \\
 BD +09 3223  &16:33:35.58&+09:06:16.3&        &  5350          &2.00     &$-2.26$   & 0.045  & 23      \\
 BD +11 2998  &16:30:16.78&+10:59:51.7&  F8    &  5373          &2.30     &$-1.36$   & 0.024  & 23      \\
 \hline
 \end{tabular}
 \caption{A portion of the Table \ref{tabla.final} is shown for guidance regarding 
its format and content. The full table is electronically available at 
http://www.ucm.es/info/Astrof/MILES/miles.html.
\label{tabla.final}}
 \end{table*}

Final stellar spectra were corrected for telluric absorptions of O$_2$
(headbands at $\sim 6280$\,\AA\ and 6870\,\AA) and H$_2$O ($\sim
7180$\,\AA) by the classic technique of dividing into a reference,
telluric spectrum. In short, an averaged, telluric spectrum for the
whole stellar library was derived from $\sim 50$ hot (O -- A types)
MILES spectra, the ones were previously shifted in the spectral
direction by cross-correlating around their telluric regions to
prevent systematic offsets among telluric lines arising from stellar,
radial velocity corrections. The resulting spectrum was continuum
normalised, with regions free from telluric absorptions
being artificially set to 1. The spectral regions for which
corrections have been carried out run in the ranges $\sim 6000 -
6320$\,AA, $\sim 6760 - 7070$\,AA\ and $\sim 7120 - 7380$\,AA. For
each star in the library, an specific, normalized telluric spectrum
matching the position of the stellar telluric features was again
derived from cross-correlation. Such a specific, normalized telluric
spectrum was used as a seed to generate a whole set of scaled,
normalized telluric spectra with different line-strengths. The stellar
spectrum was then divided into each normalized, telluric spectrum of
the set. The residuals of the corrected pixels with respect to local,
linear fits to these regions were computed separately for the O$_2$
and H$_2$O bands in each case. Finally, the corrections minimizing the
residuals for the different bands were considered as final solutions.

As pointed out in Stevenson (1994), the present technique may not be
completely optimal when, as in this case, the velocity dispersion of
the spectra is lower than $\sim 40$\,km\,s$^{-1}$. It is therefore
important to emphasize that individual measurements of line-strengths
within the corrected regions may not be totally safe. The major
improvement arises, however, when different stellar spectra are
combined together (following, for instance, the prescriptions of SSP
evolutionary synthesis models), as possible residuals coming from
uncertainties in the telluric corrections are proven to cancel because
of their different positions in the de-redshifted stellar spectrum (see
Vazdekis et al.~2006; in preparation).

 \section{Quality control}

In order to verify the reliability of our spectra and the reduction
procedure, we have: {\it (i)} carried out a detailed analysis of stars with
repeated observations to check the internal consistency, and {\it(ii)}
compared synthetic photometry on the spectra with published values.

\subsection{Internal consistency}
\begin{figure*}
\resizebox{0.8\hsize}{!}{\includegraphics[angle=-90]{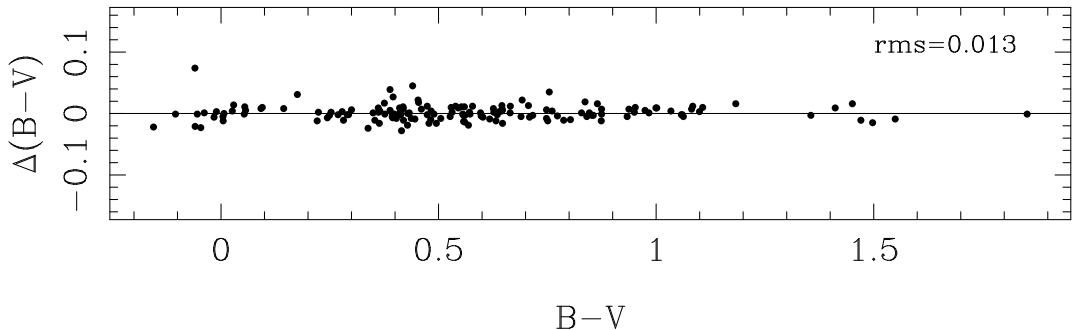}}
 \caption{Internal B-V error for our repeated observations versus
published $(B-V)$ colours from Mermilliod et al. (1997). 
 \label{fig-comparison.nights}}
  \end{figure*}

There are in total 157 repeated observations for 151 different stars in
the library with independent flux-calibrations. We have measured synthetic
$(B-V)$ colours by applying the relevant Johnson \& Morgan (1953) filter
transmission curves (photoelectric USA versions) to these fully
calibrated spectra that are not corrected for interstellar reddening and
found, from pairwise comparisons, a global RMS dispersion of 0.013
magnitudes. This is an estimate of the random errors affecting the
flux-calibration of the library. Figure \ref{fig-comparison.nights} illustrates
this comparison by plotting
offsets in $(B-V)$ colours from repeated observations versus Johnson colours from
the Lausanne database (http://obswww.unige.ch/gcpd/gcpd.html)
(Mermilliod et al. 1997). Note that this error does not account for
possible systematic errors uniformly affecting the whole library.

 \subsection{External comparisons}
\label{sec62}

\begin{figure*}
 \resizebox{0.8\hsize}{!}{\includegraphics[angle=-90]{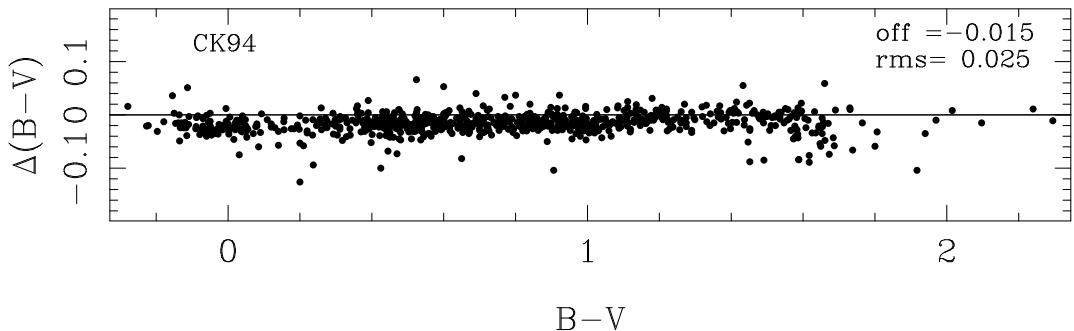}}
 \caption{Residuals of the comparison between synthetic and empirical
          $(B-V)$ colours from the Lausanne database. The numbers within
          the panels show the derived mean offsets and standard
          deviations. See the text for details.}
 \label{test.photometry}
 \end{figure*}

Once we have obtained an estimate of the random errors affecting the flux
calibration, the comparison with external measurements can provide a good
constraint of the possible systematic uncertainties affecting that
calibration. In this sense, we have carried out a comparison between the
synthetic $(B-V)$ colours derived from our library spectra and the
corresponding colours extracted from the Lausanne photometric database
(http://obswww.unige.ch/gcpd/gcpd.html)  (Mermilliod et al. 1997). The
Johnson $(B-V)$ colour has been chosen to perform such a comparison in since
it constitutes by far the largest photometric dataset in that catalogue.
In order to constrain the non-trivial problem of using an accurate zero
point for the B and V filters, we used the SED from Bohlin \& Gilliland (2004),
based on high signal to noise STIS observations and extrapolated to higher
wavelengths using Kurucz model atmospheres.
After normalising the MILES spectra to the SED of VEGA individually we
measured the synthetic $(B-V)$ colours by applying the same method
described above for the internal consistency check. Figure
\ref{test.photometry} shows the colour residuals (synthetic minus
catalogue values) versus $(B-V)$ from the catalogue for the spectrum of Vega.
The mean offsets and standard deviations of the
comparison is indicated within each panel. The absolute
value of the offset is small (around 0.015 mag), which sets an upper limit to the
systematic uncertainties of our photometry in the spectral range of MILES
up to $\sim$ 6000~\AA.

Furthermore, the measured RMS dispersions are, as expected, larger than
the previous standard deviation derived in the internal comparison, and
they can be easily understood just by assuming a typical error in the
compiled photometry of the external catalogue of $\sim$ 0.02 magnitudes.

 \section{Comparison with other spectral libraries}

Since MILES contains a considerably number of stars in common with
other libraries, it is an interesting task to study how well our
spectra compare with theirs. We carried out this test in two
steps. First, we analyzed the differences in synthetic colours
computed using the common spectral wavelength range, and, next, we compared
the measured Lick/IDS indices.

For the photometric comparison we selected the following spectral
libraries: Jones (1997), ELODIE (Prugniel \& Soubiran 2001), STELIB
(Le Borgne et al. 2003) and INDO-US (Valdes et al. 2004). 

All the libraries were broadened to match the
poorest spectral resolution (3~\AA~FWHM) of the different datasets,
and re-sampled to a common 0.9~\AA/pixel linear dispersion. 
Prior to the comparison,
we present some details of the comparing libraries:
\begin{itemize}
\item Jones (1997): This library, with 295 stars in common with MILES,
covers two narrow wavelength ranges (3856--4476~\AA\ and
4795--5465~\AA) and has been flux calibrated. However, the spectra
are not corrected from interstellar reddening.
\item ELODIE (Prugniel \& Soubiran 2001): MILES has 202 spectra in common with the dataset of
ELODIE. Their spectra cover a wavelength range from 4100 to 6800~\AA.
\item STELIB (Le Borgne et al. 2003): The MILES database contains 106 stars in common
with STELIB. This library is flux calibrated and corrected for
interstellar extinction.
\item INDO-US  (Valdes et al. 2004): 
We have analyzed 310 stars in common between this library and
MILES. The library has not
calibrated in flux. However, as it was said before, the authors have fitted
the continuum shape of each spectrum to standard SEDs from the Pickles'
(1998) library with a close match in spectral type. The library includes
14 stars for which a flat continuum applied, and which we have not used in the
comparison.
\end{itemize}

 \subsection{Photometric comparison} 

In order to carry out this comparison, we defined 7 box filters  
in different spectral regions (see the definitions in Table~\ref{tab.filtrosbox}).
We measured relative fluxes within the filters and a combination of them provided 
several (as many as 5, depending on the library) synthetic colours for the stars of the 
different datasets.

 \begin{table}
 \centering
 \begin{tabular}{l c c}
 Filter & $\lambda_{\rm c}$(\AA) & Width (\AA) \\
 \hline\hline
 b4000  &   4000                   & 200          \\
 b4300  &   4300                   & 200          \\
 b4900  &   4900                   & 200          \\
 b5300  &   5300                   & 200          \\
 b4600  &   4600                   & 800           \\
 b5400  &   5400                   & 800          \\
 b6200  &   6200                   & 800          \\
 \hline
\end{tabular}
 \caption{Characteristics of the box filters used in the comparison with other
 libraries. We list the central wavelengths ($\lambda_{\rm c}$) and the 
 filter widths.
\label{tab.filtrosbox}}
 \end{table}

\begin{figure*}
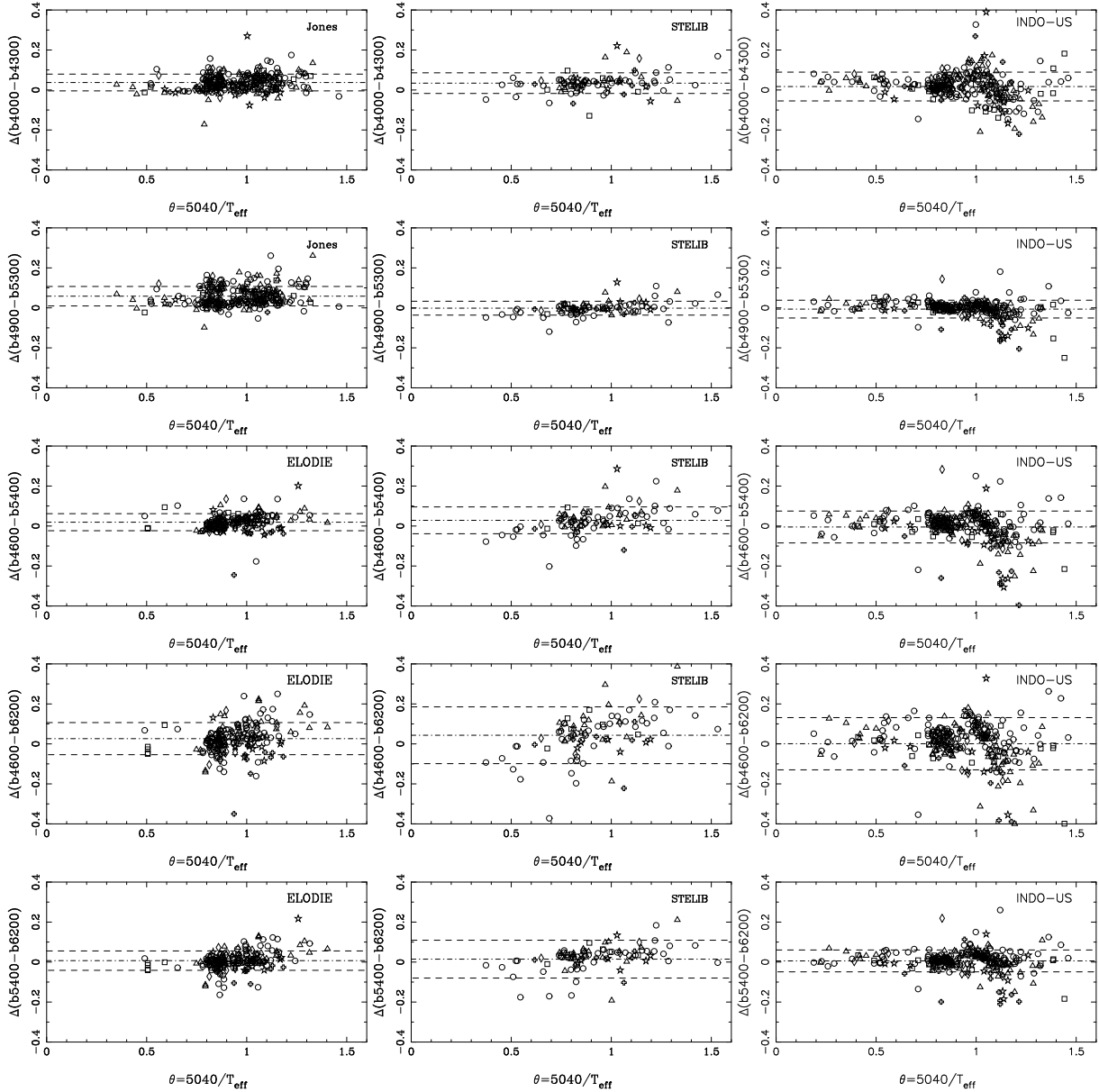

  \resizebox{0.3\hsize}{!}{\includegraphics[angle=-90]{b40.b43.jones.ps}}
 \resizebox{0.3\hsize}{!}{\includegraphics[angle=-90]{b40.b43.stelib.ps}}
  \resizebox{0.3\hsize}{!}{\includegraphics[angle=-90]{b40.b43.rose.ps}}
   \resizebox{0.3\hsize}{!}{\includegraphics[angle=-90]{b49.b53.jones.ps}}
 \resizebox{0.3\hsize}{!}{\includegraphics[angle=-90]{b49.b53.stelib.ps}}
  \resizebox{0.3\hsize}{!}{\includegraphics[angle=-90]{b49.b53.rose.ps}}
    \resizebox{0.3\hsize}{!}{\includegraphics[angle=-90]{b46.b54.elodie.ps}}
 \resizebox{0.3\hsize}{!}{\includegraphics[angle=-90]{b46.b54.stelib.ps}}
  \resizebox{0.3\hsize}{!}{\includegraphics[angle=-90]{b46.b54.rose.ps}}
     \resizebox{0.3\hsize}{!}{\includegraphics[angle=-90]{b46.b62.elodie.ps}}
 \resizebox{0.3\hsize}{!}{\includegraphics[angle=-90]{b46.b62.stelib.ps}}
  \resizebox{0.3\hsize}{!}{\includegraphics[angle=-90]{b46.b62.rose.ps}}
      \resizebox{0.3\hsize}{!}{\includegraphics[angle=-90]{b54.b62.elodie.ps}}
 \resizebox{0.3\hsize}{!}{\includegraphics[angle=-90]{b54.b62.stelib.ps}}
  \resizebox{0.3\hsize}{!}{\includegraphics[angle=-90]{b54.b62.rose.ps}}
 \caption{Residuals of the comparison between synthetic colours measured in
 MILES and in other stellar libraries versus  $\theta=5040/T_{\rm eff}$. The different
 symbols indicate stars of different metallicities, as coded in
 Fig. \ref{fig-hrdiagram}. Dashed lines correspond to  $1\sigma$ RMS.
\label{fig-b40.b43}}
 \end{figure*}

Fig. \ref{fig-b40.b43} shows the residuals of the synthetic colours
measured on MILES and on the spectra from the other datasets. The
residuals have been obtained as the colour in the comparing library
minus the colour in MILES. Table~\ref{table.comparafotometria} lists
the mean offsets and the dispersions found in these comparisons.  The
agreement is generally good, with the systematic effect in all the
cases lower than 0.06 mag. Note that, when significant, the offsets
are generally in the sense of MILES being somewhat bluer than the
previous libraries.

\begin{figure}
 \resizebox{1.0\hsize}{!}{\includegraphics[angle=-90]{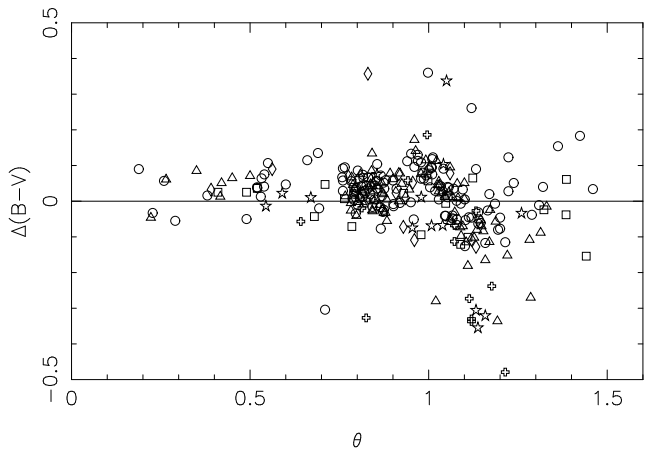}}   
  \caption{Residuals of the comparison between synthetic $(B-V)$ colours from the INDO-US library 
and MILES ($\Delta(B-V)=(B-V)_{\rm INDO-US}-(B-V)_{\rm MILES}$). Symbols are the same as in Fig.
\ref{fig-hrdiagram}.
\label{bvindo}}
\end{figure}

In general, the best agreement is obtained with INDO-US library, for
which we do not find significant differences in the broader colours
(lower part of Table~\ref{table.comparafotometria}). To further
explore the possible differences in the photometric calibration, we
have also measured synthetic $(B-V)$ colours on all INDO-US stars in
common with MILES, obtaining a mean offset between both libraries of
$\Delta(B-V)=0.000$ mag, with a RMS dispersion of $0.102$ mag (see
Figure~\ref{bvindo}). This gives us confidence about our photometric
calibration, since Pickles' library (Pickles 1998) is considered to be
very well flux calibrated.  However, in both  Figs~\ref{fig-b40.b43} and
\ref{bvindo}, residuals seems to follow a different behaviour for
stars colder and hotter than $\sim$5600 K ($\theta=0.9$). Stars colder
than this temperature exhibit larger residuals, with some of them as
high as 0.4 magnitudes. This could be due to the different criteria
applied by these authors to the cool and hot stars with the aim of assigning a continuum
shape to each star from the Pickles library.  To examine these
differences in more detail, we have compared the INDO-US spectra of
the stars with highest residuals with those from MILES, ELODIE or
STELIB libraries, finding some stars for which INDO-US provide very different
continuum shapes. Therefore, although in general the shape of the
continuum for the stars of this library has been well approximated, the
method applied by the authors can lead to some large errors in the
assigned shape of the continuum of some cool stars.

\begin{figure}
 \resizebox{1.0\hsize}{!}{\includegraphics[angle=-90]{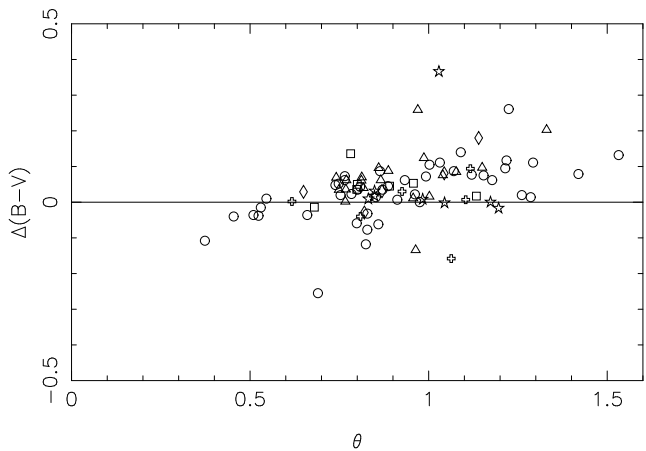}}   
  \caption{Residuals of the comparison between synthetic $(B-V)$ colours from the STELIB dataset 
and MILES ($\Delta(B-V)=(B-V)_{\rm STELIB}-(B-V)_{\rm MILES}$). Symbols are the same as in Fig.
\ref{fig-hrdiagram}.
\label{bvstelib}}
\end{figure}

Concerning the comparison with the STELIB dataset, and in order to
quantify the differences in the photometric calibration, we have also
compared the synthetic $(B-V)$ colours obtained in this library with
those measured in the MILES spectra.  The residuals, plotted in
Figure~\ref{bvstelib}, reveal that, on the average, STELIB spectra are
redder than MILES ones by $\Delta(B-V)=0.010$ mag, with a RMS
dispersion of $0.100$ mag.  It is interesting to compare this
dispersion with the one obtained in the comparison of MILES with
tabulated colours from the Lausanne database (section~\ref{sec62}; RMS
$\simeq 0.024$ mag). This suggests that most of the above dispersion comes
from uncertainties in the calibration of STELIB. Le Borgne et
al. (2003) indeed find, in their own comparison with the Lausanne
data), an RMS dispersion of 0.083 mag.

The largest systematic differences in the comparison of the colours
between MILES and other libraries are obtained with the Jones'
dataset. One of the reasons for these discrepancies could be the
absence of interstellar reddening correction in Jones' library.  In
order to test this, we have searched for a possible correlation between
the residuals of the ($b4900-b5300$) colour and the colour excesses
$E(B-V)$ of the stars in the comparison. The results show that,
although the three stars with the highest $E(B-V)$ values are also the
ones with the highest colour differences, for the rest of the stars
there does not exist such a correlation. Therefore, we do not know the
causes of the reported differences, although it must be noted that the
flux calibration for some stars in Jones' library has errors higher
than 25\%, due to the selective flux losses in the spectrograph slit
(see Vazdekis 1999).

\begin{table}
 \begin{tabular}{l r r r r r r }\hline
              & \multicolumn{2}{c}{Jones}&\multicolumn{2}{c}{STELIB}&\multicolumn{2}{c}{INDO-US}\\
              & offset  & RMS   & offset & RMS   & offset & RMS\\ 
\hline\hline
 b4000-b4300  & {\bf 0.037}& 0.042 & {\bf 0.034}& 0.052& {\bf 0.017}& 0.072\\
 b4900-b5300  & {\bf 0.058}& 0.048 &$-$0.001& 0.035&{\bf $-$0.006}& 0.044\\ 
 \hline       
              &\multicolumn{2}{c}{ELODIE}&\multicolumn{2}{c}{STELIB}&\multicolumn{2}{c}{INDO-US}\\
              &offset   &RMS    & offset  &RMS   & offset  & RMS\\
\hline\hline
 b4600-b5400  & {\bf 0.019}&0.042  &{\bf 0.028}&0.067  &$-$0.005&0.078\\ 
 b4600-b6200  & {\bf 0.007}&0.048  &{\bf 0.043}&0.142  & 0.001&0.131\\ 
 b5400-b6200  & {\bf 0.026}&0.080  &0.014&0.095  & 0.005&0.055\\
\hline
 \end{tabular}
 \caption{Mean offsets and standard deviations (RMS) from the comparison 
 between the synthetic colours of MILES and other libraries. Bold typeface is used when
the offsets are statistically different from 0 for a 95\% level of confidence. 
\label{table.comparafotometria}}
 \end{table}

 \subsection{Comparison of the Lick indices}
  
\begin{figure*}
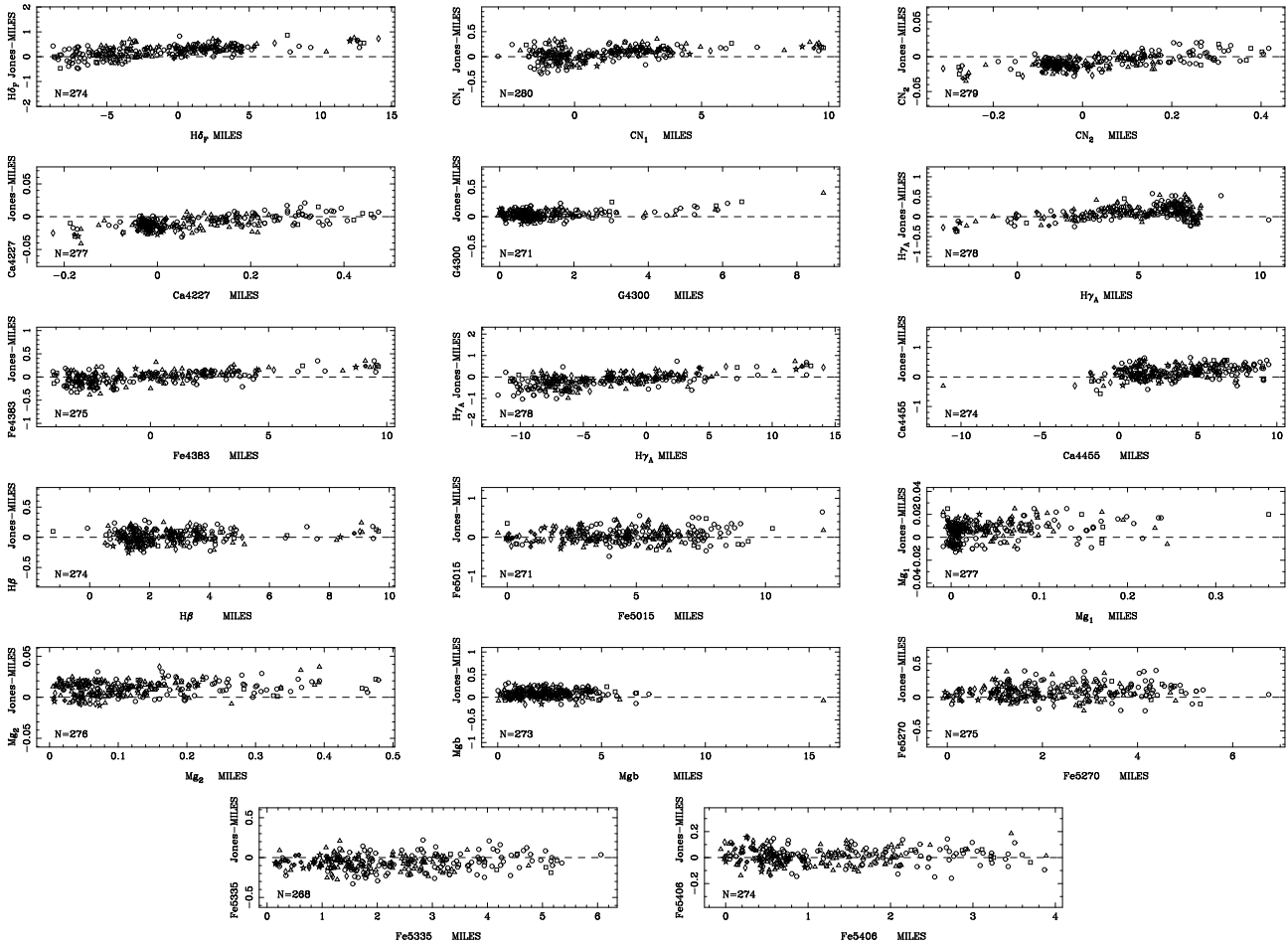

 
\resizebox{0.30\hsize}{!}{\includegraphics[bbllx=297,bblly=84,bburx=475,bbury=755,angle=-90]{hda.jones.new4.ps}}
\hspace{0.5cm}
\resizebox{0.30\hsize}{!}{\includegraphics[bbllx=297,bblly=84,bburx=475,bbury=755,angle=-90]{hdf.jones.new4.ps}}
\hspace{0.5cm}
\resizebox{0.30\hsize}{!}{\includegraphics[bbllx=297,bblly=84,bburx=475,bbury=755,angle=-90]{cn1.jones.new4.ps}}

\vspace{0.7cm}

\resizebox{0.30\hsize}{!}{\includegraphics[bbllx=297,bblly=84,bburx=475,bbury=755,angle=-90]{cn2.jones.new4.ps}}
\hspace{0.5cm}
\resizebox{0.30\hsize}{!}{\includegraphics[bbllx=297,bblly=84,bburx=475,bbury=755,angle=-90]{ca4227.jones.new4.ps}}
\hspace{0.5cm}
\resizebox{0.30\hsize}{!}{\includegraphics[bbllx=297,bblly=84,bburx=475,bbury=755,angle=-90]{g4300.jones.new4.ps}}

\vspace{0.7cm}

\resizebox{0.30\hsize}{!}{\includegraphics[bbllx=297,bblly=84,bburx=475,bbury=755,angle=-90]{hga.jones.new4.ps}}
\hspace{0.5cm}
\resizebox{0.30\hsize}{!}{\includegraphics[bbllx=297,bblly=84,bburx=475,bbury=755,angle=-90]{hgf.jones.new4.ps}}
\hspace{0.5cm}
\resizebox{0.30\hsize}{!}{\includegraphics[bbllx=297,bblly=84,bburx=475,bbury=755,angle=-90]{fe4383.jones.new4.ps}}

\vspace{0.7cm}

\resizebox{0.30\hsize}{!}{\includegraphics[bbllx=297,bblly=84,bburx=475,bbury=755,angle=-90]{hbeta.jones.new4.ps}}
\hspace{0.5cm}
\resizebox{0.30\hsize}{!}{\includegraphics[bbllx=297,bblly=84,bburx=475,bbury=755,angle=-90]{fe5015.jones.new4.ps}}
\hspace{0.5cm}
\resizebox{0.30\hsize}{!}{\includegraphics[bbllx=297,bblly=84,bburx=475,bbury=755,angle=-90]{mg1.jones.new4.ps}}

\vspace{0.7cm}

\resizebox{0.30\hsize}{!}{\includegraphics[bbllx=297,bblly=84,bburx=475,bbury=755,angle=-90]{mg2.jones.new4.ps}}
\hspace{0.5cm}
\resizebox{0.30\hsize}{!}{\includegraphics[bbllx=297,bblly=84,bburx=475,bbury=755,angle=-90]{mgb.jones.new4.ps}}
\hspace{0.5cm}
\resizebox{0.30\hsize}{!}{\includegraphics[bbllx=297,bblly=84,bburx=475,bbury=755,angle=-90]{fe5270.jones.new4.ps}}

\vspace{0.7cm}

\resizebox{0.30\hsize}{!}{\includegraphics[bbllx=297,bblly=84,bburx=475,bbury=755,angle=-90]{fe5335.jones.new4.ps}}
\hspace{0.5cm}
\resizebox{0.30\hsize}{!}{\includegraphics[bbllx=297,bblly=84,bburx=475,bbury=755,angle=-90]{fe5406.jones.new4.ps}} 

\vspace{0.7cm}

 \caption{Differences of the Lick/IDS indices between MILES spectra and  Jones'
 library against the indices measured in MILES.
 The number of stars in the comparison are displayed within each panel.
The meaning of the symbols is the same as in Figure\ref{fig-hrdiagram}. 
\label{fig.jones.lickindices}}
 \end{figure*}

\begin{figure*}
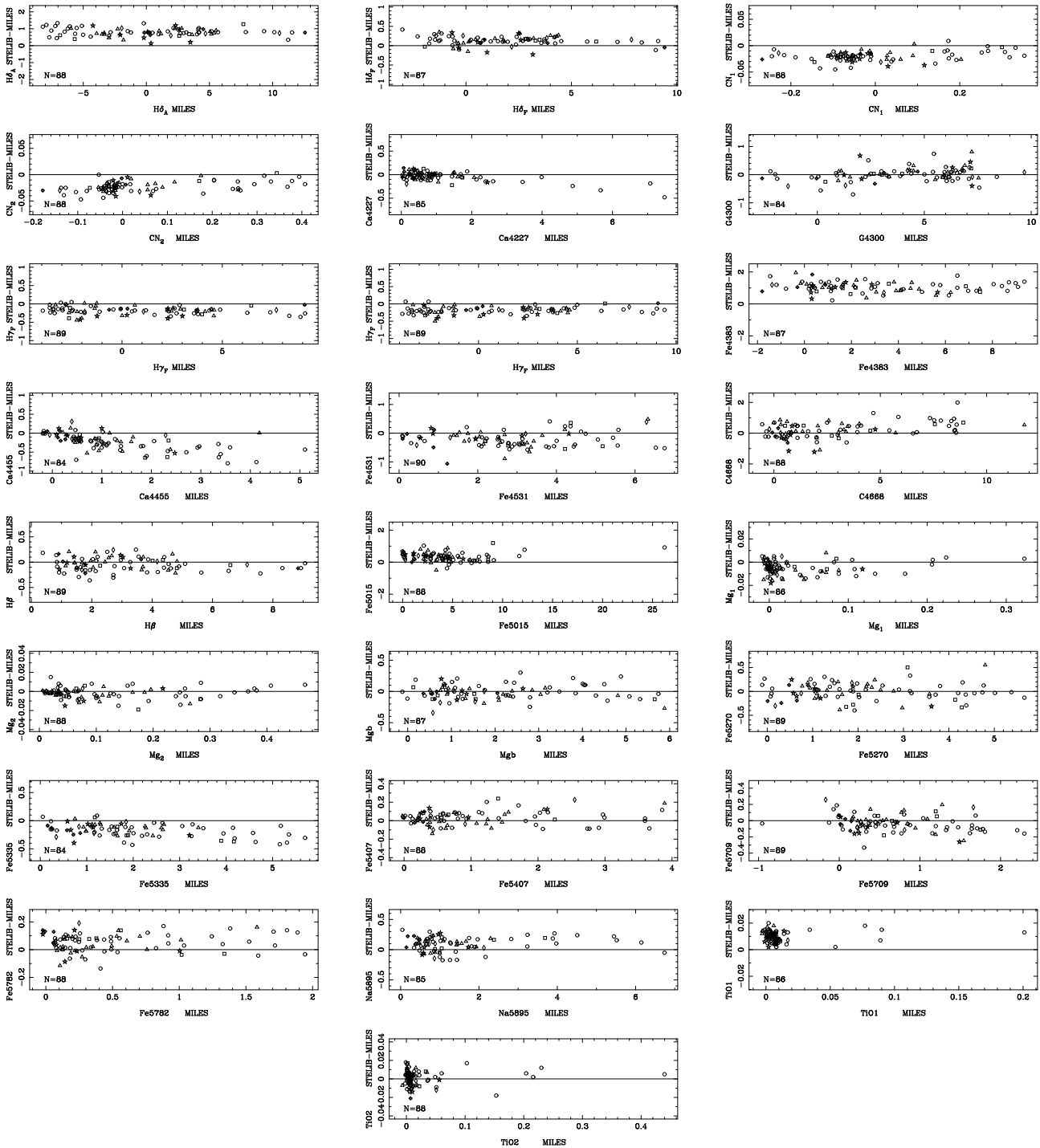

\resizebox{0.30\hsize}{!}{\includegraphics[bbllx=297,bblly=84,bburx=475,bbury=757,angle=-90]{hda.stelib.new4.ps}}
\hspace{0.5cm}
\resizebox{0.30\hsize}{!}{\includegraphics[bbllx=297,bblly=84,bburx=475,bbury=757,angle=-90]{hdf.stelib.new4.ps}}  
\hspace{0.5cm}
\resizebox{0.30\hsize}{!}{\includegraphics[bbllx=297,bblly=84,bburx=475,bbury=757,angle=-90]{cn1.stelib.new4.ps}}  

\vspace{0.7cm}

\resizebox{0.30\hsize}{!}{\includegraphics[bbllx=297,bblly=84,bburx=475,bbury=757,angle=-90]{cn2.stelib.new4.ps}}  
\hspace{0.5cm}
\resizebox{0.30\hsize}{!}{\includegraphics[bbllx=297,bblly=84,bburx=475,bbury=757,angle=-90]{ca4227.stelib.new4.ps}}
\hspace{0.5cm}
\resizebox{0.30\hsize}{!}{\includegraphics[bbllx=297,bblly=84,bburx=475,bbury=757,angle=-90]{g4300.stelib.new4.ps}}  

\vspace{0.7cm}

\resizebox{0.30\hsize}{!}{\includegraphics[bbllx=297,bblly=84,bburx=475,bbury=757,angle=-90]{hga.stelib.new4.ps}}  
\hspace{0.5cm}
\resizebox{0.30\hsize}{!}{\includegraphics[bbllx=297,bblly=84,bburx=475,bbury=757,angle=-90]{hgf.stelib.new4.ps}}  
\hspace{0.5cm}
\resizebox{0.30\hsize}{!}{\includegraphics[bbllx=297,bblly=84,bburx=475,bbury=757,angle=-90]{fe4383.stelib.new4.ps}}

\vspace{0.7cm}

\resizebox{0.30\hsize}{!}{\includegraphics[bbllx=297,bblly=84,bburx=475,bbury=757,angle=-90]{ca4455.stelib.new4.ps}}  
\hspace{0.5cm}
\resizebox{0.30\hsize}{!}{\includegraphics[bbllx=297,bblly=84,bburx=475,bbury=757,angle=-90]{fe4531.stelib.new4.ps}}  
\hspace{0.5cm}
\resizebox{0.30\hsize}{!}{\includegraphics[bbllx=297,bblly=84,bburx=475,bbury=757,angle=-90]{fe4668.stelib.new4.ps}}

\vspace{0.7cm}

\resizebox{0.30\hsize}{!}{\includegraphics[bbllx=297,bblly=84,bburx=475,bbury=757,angle=-90]{hbeta.stelib.new4.ps}}
\hspace{0.5cm}
\resizebox{0.30\hsize}{!}{\includegraphics[bbllx=297,bblly=84,bburx=475,bbury=757,angle=-90]{fe5015.stelib.new4.ps}}
\hspace{0.5cm}
\resizebox{0.30\hsize}{!}{\includegraphics[bbllx=297,bblly=84,bburx=475,bbury=757,angle=-90]{mg1.stelib.new4.ps}}

\vspace{0.7cm}

\resizebox{0.30\hsize}{!}{\includegraphics[bbllx=297,bblly=84,bburx=475,bbury=757,angle=-90]{mg2.stelib.new4.ps}}
\hspace{0.5cm}
\resizebox{0.30\hsize}{!}{\includegraphics[bbllx=297,bblly=84,bburx=475,bbury=757,angle=-90]{mgb.stelib.new4.ps}}
\hspace{0.5cm}
\resizebox{0.30\hsize}{!}{\includegraphics[bbllx=297,bblly=84,bburx=475,bbury=757,angle=-90]{fe5270.stelib.new4.ps}}

\vspace{0.7cm}

\resizebox{0.30\hsize}{!}{\includegraphics[bbllx=297,bblly=84,bburx=475,bbury=757,angle=-90]{fe5335.stelib.new4.ps}}
\hspace{0.5cm}
\resizebox{0.30\hsize}{!}{\includegraphics[bbllx=297,bblly=84,bburx=475,bbury=757,angle=-90]{fe5406.stelib.new4.ps}}
\hspace{0.5cm}
\resizebox{0.30\hsize}{!}{\includegraphics[bbllx=297,bblly=84,bburx=475,bbury=757,angle=-90]{fe5709.stelib.new4.ps}}

\vspace{0.7cm}

\resizebox{0.30\hsize}{!}{\includegraphics[bbllx=297,bblly=84,bburx=475,bbury=757,angle=-90]{fe5782.stelib.new4.ps}}
\hspace{0.5cm}
\resizebox{0.30\hsize}{!}{\includegraphics[bbllx=297,bblly=84,bburx=475,bbury=757,angle=-90]{na5895.stelib.new4.ps}}
\hspace{0.5cm}
\resizebox{0.30\hsize}{!}{\includegraphics[bbllx=297,bblly=84,bburx=475,bbury=757,angle=-90]{tio1.stelib.new4.ps}}

\vspace{0.7cm}

\resizebox{0.30\hsize}{!}{\includegraphics[bbllx=297,bblly=84,bburx=475,bbury=757,angle=-90]{tio2.stelib.new4.ps}}

\vspace{0.7cm}

\caption{Differences between the Lick/IDS indices measured in 
MILES and STELIB spectra against MILES's indices. 
  In each panel, the number of stars in the comparison is 
  indicated.
 Stars with different metallicities
 are displayed with different symbols as in Figure \ref{fig-hrdiagram}. 
\label{fig.stelib.lickindices}}
\end{figure*}

\begin{figure*}
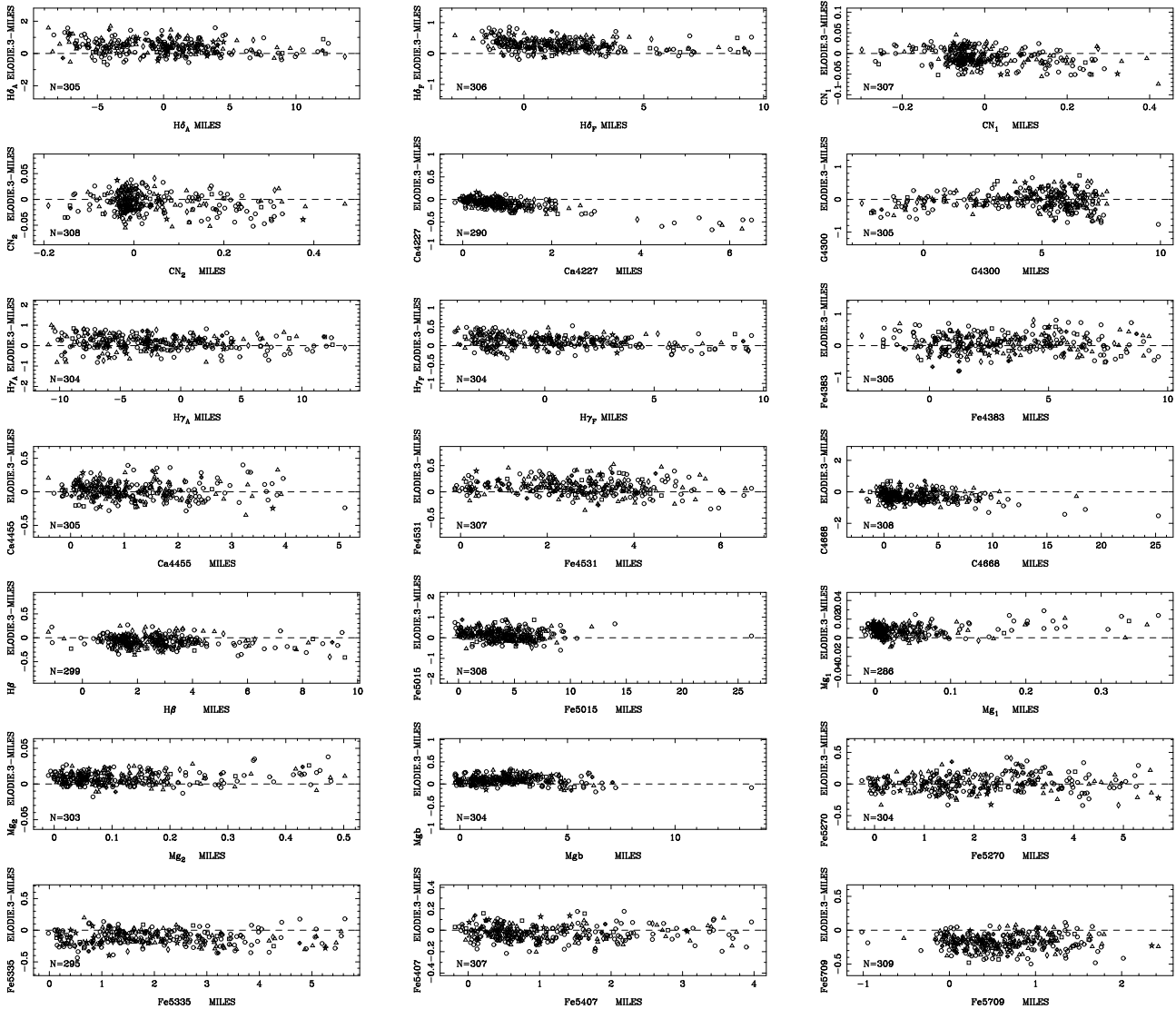


\resizebox{0.30\hsize}{!}{\includegraphics[bbllx=297,bblly=84,bburx=475,bbury=757,angle=-90]{hda.elodie.new4.ps}}
\hspace{0.5cm}
\resizebox{0.30\hsize}{!}{\includegraphics[bbllx=297,bblly=84,bburx=475,bbury=757,angle=-90]{hdf.elodie.new4.ps}}
\hspace{0.5cm}
\resizebox{0.30\hsize}{!}{\includegraphics[bbllx=297,bblly=84,bburx=475,bbury=757,angle=-90]{cn1.elodie.new4.ps}}

\vspace{0.7cm}

\resizebox{0.30\hsize}{!}{\includegraphics[bbllx=297,bblly=84,bburx=475,bbury=757,angle=-90]{cn2.elodie.new4.ps}}
\hspace{0.5cm}
\resizebox{0.30\hsize}{!}{\includegraphics[bbllx=297,bblly=84,bburx=475,bbury=757,angle=-90]{ca4227.elodie.new4.ps}}
\hspace{0.5cm}
\resizebox{0.30\hsize}{!}{\includegraphics[bbllx=297,bblly=84,bburx=475,bbury=757,angle=-90]{g4300.elodie.new4.ps}}

\vspace{0.7cm}

\resizebox{0.30\hsize}{!}{\includegraphics[bbllx=297,bblly=84,bburx=475,bbury=757,angle=-90]{hga.elodie.new4.ps}}
\hspace{0.5cm}
\resizebox{0.30\hsize}{!}{\includegraphics[bbllx=297,bblly=84,bburx=475,bbury=757,angle=-90]{hgf.elodie.new4.ps}}
\hspace{0.5cm}
\resizebox{0.30\hsize}{!}{\includegraphics[bbllx=297,bblly=84,bburx=475,bbury=757,angle=-90]{fe4383.elodie.new4.ps}}

\vspace{0.7cm}

\resizebox{0.30\hsize}{!}{\includegraphics[bbllx=297,bblly=84,bburx=475,bbury=757,angle=-90]{ca4455.elodie.new4.ps}}
\hspace{0.5cm}
\resizebox{0.30\hsize}{!}{\includegraphics[bbllx=297,bblly=84,bburx=475,bbury=757,angle=-90]{fe4531.elodie.new4.ps}}
\hspace{0.5cm}
\resizebox{0.30\hsize}{!}{\includegraphics[bbllx=297,bblly=84,bburx=475,bbury=757,angle=-90]{fe4668.elodie.new4.ps}}

\vspace{0.7cm}

\resizebox{0.30\hsize}{!}{\includegraphics[bbllx=297,bblly=84,bburx=475,bbury=757,angle=-90]{hbeta.elodie.new4.ps}}
\hspace{0.5cm}
\resizebox{0.30\hsize}{!}{\includegraphics[bbllx=297,bblly=84,bburx=475,bbury=757,angle=-90]{fe5015.elodie.new4.ps}}
\hspace{0.5cm}
\resizebox{0.30\hsize}{!}{\includegraphics[bbllx=297,bblly=84,bburx=475,bbury=757,angle=-90]{mg1.elodie.new4.ps}}

\vspace{0.7cm}

\resizebox{0.30\hsize}{!}{\includegraphics[bbllx=297,bblly=84,bburx=475,bbury=757,angle=-90]{mg2.elodie.new4.ps}}
\hspace{0.5cm}
\resizebox{0.30\hsize}{!}{\includegraphics[bbllx=297,bblly=84,bburx=475,bbury=757,angle=-90]{mgb.elodie.new4.ps}}
\hspace{0.5cm}
\resizebox{0.30\hsize}{!}{\includegraphics[bbllx=297,bblly=84,bburx=475,bbury=757,angle=-90]{fe5270.elodie.new4.ps}}

\vspace{0.7cm}

\resizebox{0.30\hsize}{!}{\includegraphics[bbllx=297,bblly=84,bburx=475,bbury=757,angle=-90]{fe5335.elodie.new4.ps}}
\hspace{0.5cm}
\resizebox{0.30\hsize}{!}{\includegraphics[bbllx=297,bblly=84,bburx=475,bbury=757,angle=-90]{fe5406.elodie.new4.ps}}
\hspace{0.5cm}
\resizebox{0.30\hsize}{!}{\includegraphics[bbllx=297,bblly=84,bburx=475,bbury=757,angle=-90]{fe5709.elodie.new4.ps}}

 \vspace{0.7cm}

\resizebox{0.30\hsize}{!}{\includegraphics[bbllx=297,bblly=84,bburx=475,bbury=757,angle=-90]{fe5782.elodie.new3.ps}}

 \vspace{0.7cm}

 \caption{Differences between  the Lick indices measured on MILES and in the most recent version of ELODIE
  spectra against MILES's indices.  
  In each panel, the number of stars fitted and the 
 slope of the fit are indicated. Stars with different metallicities
 are displayed with different symbols as in Figure \ref{fig-hrdiagram}. 
\label{fig.elodie.lickindices}}
 \end{figure*}

In this section we present the comparison between the Lick/IDS indices
measured on the common stars between MILES and the other stellar
libraries. We only show the results for the libraries whose spectra
have been incorporated into stellar population models (i.e. Jones'
library, STELIB and ELODIE.3).  Figures~\ref{fig.jones.lickindices},
\ref{fig.stelib.lickindices} and \ref{fig.elodie.lickindices} show this comparison. For each index
we have fitted a straight line and the slope is indicated within each
panel. These fits have been obtained iteratively, by removing, in each
step, the stars that deviated more than 3 $\sigma$. The final numbers
of stars are also given in the panels. Table~\ref{table.lick.todas}
lists the coefficients of the fits together with the corresponding
RMS.  
As can be seen, the slopes are, in general, around one, except 
in the comparison with Jones's, where the slopes are always smaller
than 1, which means that there exists a general trend for our
strongest indices to be weaker than in this library.

There exists also small differences in the intercept of the fits. 
It is important to remark that these systematic effects should be
taken into account when comparing predictions of models using
different spectral libraries. 
As a check observers who want to
compare galaxy observations with models could include several
MILES stars in their observing run in order to check that there are no
systematic differences in the line-strengths. Such a check would of course apply
for people using other libraries, like STELIB. We emphasize however that this is
not strictly necessary, because the MILES database has been fully 
flux calibrated with the explicit purpose of making observations of line
strengths easy and reproducible.
  
  \begin{table*}
 \centering
  \begin{tabular}{l r r r r r r r r r }
              & \multicolumn{3}{c}{Jones}&\multicolumn{3}{c}{STELIB}&\multicolumn{3}{c}{ELODIE.3}\\
             & slope & \multicolumn{1}{c}{$a_0$}  & RMS  & slope &  \multicolumn{1}{c}{$a_0$} & RMS & slope & \multicolumn{1}{c}{$a_0$}  & RMS \\
\hline	               	  
 H$\delta_A$ &0.969& $ 0.217$ &0.269  &1.003 &$ 0.821$ &0.489 & 1.015 & $ 0.394$& 0.465 \\
 H$\delta_F$ &0.975& $ 0.059$ &0.140  &1.005 &$ 0.104$ &0.166 & 1.020 & $ 0.283$& 0.202 \\
 CN$_1$      &0.956& $-0.008$ &0.010  &0.977 &$-0.020$ &0.009 & 1.031 & $-0.009$& 0.019 \\
 CN$_2$      &0.958& $-0.008$ &0.010  &0.966 &$-0.023$ &0.012 & 1.001 & $-0.006$& 0.019\\ 
 Ca4227      &0.976& $ 0.031$ &0.062  &1.046 &$-0.031$ &0.082 & 1.092 & $-0.075$& 0.104 \\
 G4300       &0.965& $ 0.098$ &0.182  &0.971 &$-0.007$ &0.198 & 0.982 & $-0.040$& 0.276 \\
 H$\gamma_A$ &0.962& $-0.155$ &0.332  &0.967 &$-0.680$ &0.635 & 1.005 & $ 0.137$& 0.362 \\
 H$\gamma_F$ &0.975& $-0.004$ &0.138  &1.005 &$-0.183$ &0.110 & 1.012 & $ 0.106$& 0.198 \\
 Fe4383      &0.968& $ 0.160$ &0.232  &1.000 &$ 1.078$ &0.514 & 0.993 & $ 0.071$& 0.295 \\
 Ca4455      &     &     &            &1.097 &$-0.230$ &0.221 & 0.991 & $ 0.012$& 0.124 \\    
 Fe4531      &     &     &            &0.969 &$-0.209$ &0.260 & 1.002 & $ 0.088$& 0.154 \\
 C4668       &     &     &            &0.896 &$ 0.147$ &0.566 & 1.020 & $-0.268$& 0.345 \\
 H$\beta$    &0.987&$ 0.011$  &0.106  &1.005 &$-0.041$ &0.135 & 1.009 & $-0.074$& 0.118 \\
 Fe5015      &0.983&$ 0.043$  &0.175  &0.988 &$ 0.325$ &0.281 & 1.003 & $ 0.138$& 0.271 \\ 
 Mg$_1$      &0.952&$ 0.006$  &0.009  &0.980 &$-0.005$ &0.007 & 0.971 & $ 0.007$& 0.007 \\
 Mg$_2$      &0.975&$ 0.011$  &0.009  &0.996 &$-0.002$ &0.006 & 0.985 & $ 0.008$& 0.008 \\
 Mgb         &0.999&$ 0.075$  &0.095  &0.995 &$-0.015$ &0.117 & 1.005 & $ 0.083$& 0.103 \\
 Fe5270      &0.984&$ 0.080$  &0.119  &0.988 &$ 0.003$ &0.187 & 0.984 & $ 0.010$& 0.131 \\
 Fe5335      &0.986&$-0.064$  &0.101  &1.030 &$-0.134$ &0.183 & 0.998 & $-0.117$& 0.130 \\
 Fe5406      &0.994&$ 0.008$  &0.063  &0.988 &$ 0.030$ &0.075 & 0.996 & $-0.026$& 0.072 \\  
 Fe5709      &      &     &           &1.016 &$-0.032$ &0.098 & 0.955 & $-0.178$& 0.125 \\
 Fe5782      &      &     &           &0.973 &$ 0.054$ &0.069 & 0.950 & $-0.093$& 0.112 \\
 Na5849      &      &     &           &0.999 &$ 0.060$ &0.140 &       &       &         \\
 TiO$_1$     &      &     &           &0.965 &$ 0.009$ &0.005 &       &       &         \\ 
 TiO$_2$     &      &     &           &0.974 &$ 0.001$ &0.008 &       &       &         \\
\hline
 \end{tabular}
 \caption{Comparison of the Lick indices measured on MILES spectra with indices 
measured, after the corresponding spectral resolution correction, in Jones,
STELIB and ELODIE.3 libraries.
For each dataset, the first two columns list the slope and intercept (Other
library-MILES) of a straight line fit
to data in Figures \ref{fig.jones.lickindices} and \ref{fig.stelib.lickindices}.
\label{table.lick.todas}} 
\end{table*}

\section{Summary}

We have presented a new stellar library, MILES, which contains 995
stars covering the spectral range $\lambda\lambda$3500-7500~\AA\ at a
resolution of 2.3~\AA\ (FWHM). The main motivation of this work was to
provide a homogeneous set of stellar spectra to be incorporated into
population synthesis models. For this reason, special care has been
put in the homogeneity of the spectra and in the sample
selection. However, the library can also be useful for a variety of
astronomical purposes, from automatic stellar classification (Kurtz
1984) (e.g., to train neural networks) to test synthetic stellar
spectra, among others.
  
The main improvements
 with respect to other previous libraries are:
\begin{enumerate}
\item The number of stars of the sample. 
\item The homogeneity of the whole catalogue. All the stars have been 
observed with the same instrumental configuration and all the spectra share exactly the same 
wavelength scale and spectral resolution. 
\item The moderately high spectral resolution. This will allow to define new 
line-strength indices with an improved sensitivity to the stellar
population parameters, which, in turn, will help to break the
well-known degeneracies in the spectra of relatively old stellar
populations, like the one between age and metallicity.
\item The much improved  stellar parameter coverage (see Figure~\ref{fig-comparison}). 
The sample has been carefully selected to cover important regions of
the parameter space in order to provide reliable predictions for the
more critical phases of the stellar evolution.
\item The accuracy of the (relative) flux calibration. The spectra are very close to a 
true spectrophotometric system. This will allow to make predictions of
whole spectral energy distributions, and not only of the strength of
selected spectral features. The approach to compare model predictions
with galaxy spectra will be to smooth the synthetic spectra to the
same resolution as the observations, allowing us to analyze the
observed spectrum in its own system and to use all the information
contained in the data at its original spectral resolution.
\end{enumerate}

\section*{Acknowledgments}
We are in debt with the non-anonymous referee, Guy Worthey, for 
noticing the presence of scattered light on the first version 
of the library. 
We would also like to thank Scott C. Trager for the careful reading of the manuscript
and for his very useful comments and to Ricardo Schiavon for his 
advices to correct the stars for atmospheric extinction.
The INT is operated on the island of La Palma by the Royal Greenwich
Observatory at the Observatorio del  Roque de los Muchachos of the Instituto de
Astrof\'{\i}sica de Canarias. This work was supported by the  Spanish research
project AYA2003-01840. This work has made extensive use of the SIMBAD database.
We are grateful to the ASTRON-PC and CAT for generous allocation of telescope
time.

 \include{apendice}

\appendix
\section{Computing colour excesses}
\label{apendice}
For an important fraction of the stars with unknown reddenings from the
literature it has been posible to derive colour excesses by measuring synthetic
colours in the spectra. In this sense, we have employed the subset of stellar
spectra with published $E(B-V)$ values to check the accuracy of our own
determinations of reddenings. Once the method has shown its capabilities to
provide accurate colour excesses for a given interval in effective
temperature and gravity (for the whole metallicity range), we have employed the
same technique to derive $E(B-V)$ measurements for the subset of stars in those
ranges with unpublished reddenings.

\begin{figure}
\resizebox{1.0\hsize}{!}{\includegraphics{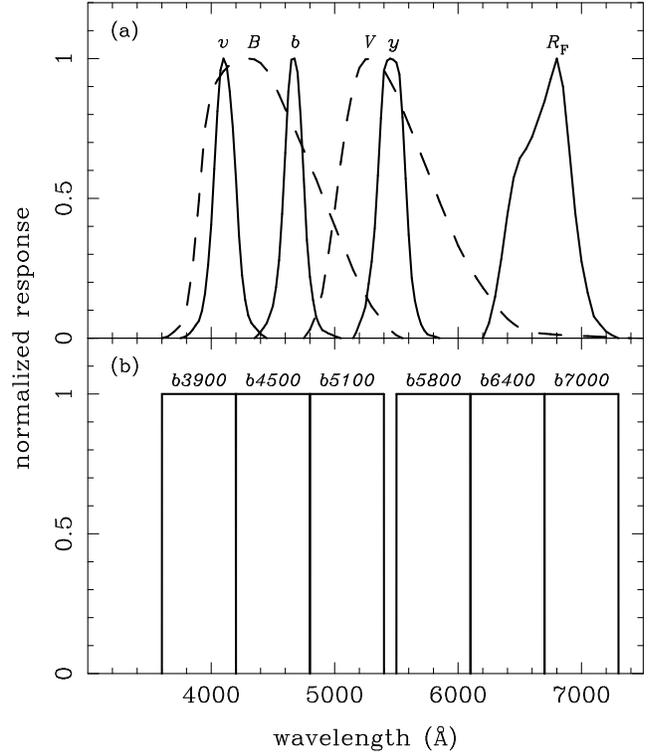}}
\caption{Normalized response functions of typical filters ---panel~(a)--- and
special filters ---panel~(b)---, employed to estimate colour excesses in the
subset of stellar spectra without published reddenings. The displayed filters
are the Johnson's $B$ and $V$, the Str\"{o}mgren's $v$, $b$ and $y$, the
Couch-Newell's $R_{\rm F}$, and box functions of 600~\AA\ width centered at
3900, 4500, 5100, 5800, 6400 and 7000~\AA.
\label{figure_filters_ebvs}}
\end{figure}

In more detail, the procedure followed to compute colour excesses has been the
following. In a first step, we identified the subset of stars with
published $E(B-V)$ values that exhibit, in a practical sense, negligible
reddenings, more precisely those verifiying $E(B-V) < 0.001$~mag. Next, we
employed the transmission curves of typical filters in the spectral range
covered by our stellar library ($\lambda\lambda$~3500--7500~\AA), to measure
synthetic colours. In particular, we used the $B$ and $V$ Johnson
filters, the Str\"{o}mgren $b$, $v$ and $y$, and the Couch-Newell $R_{\rm
F}$ filter. The normalized transmission curves of these filters are displayed
in Fig.~\ref{figure_filters_ebvs}a. In addition to those well known filters, we
have also employed a  set of filters defined as simple box functions
of 600~\AA\ width, which are also graphically shown in
Fig.~\ref{figure_filters_ebvs}b. 

\begin{figure}
\resizebox{1.0\hsize}{!}{%
\includegraphics[bb=72 169 554 773]{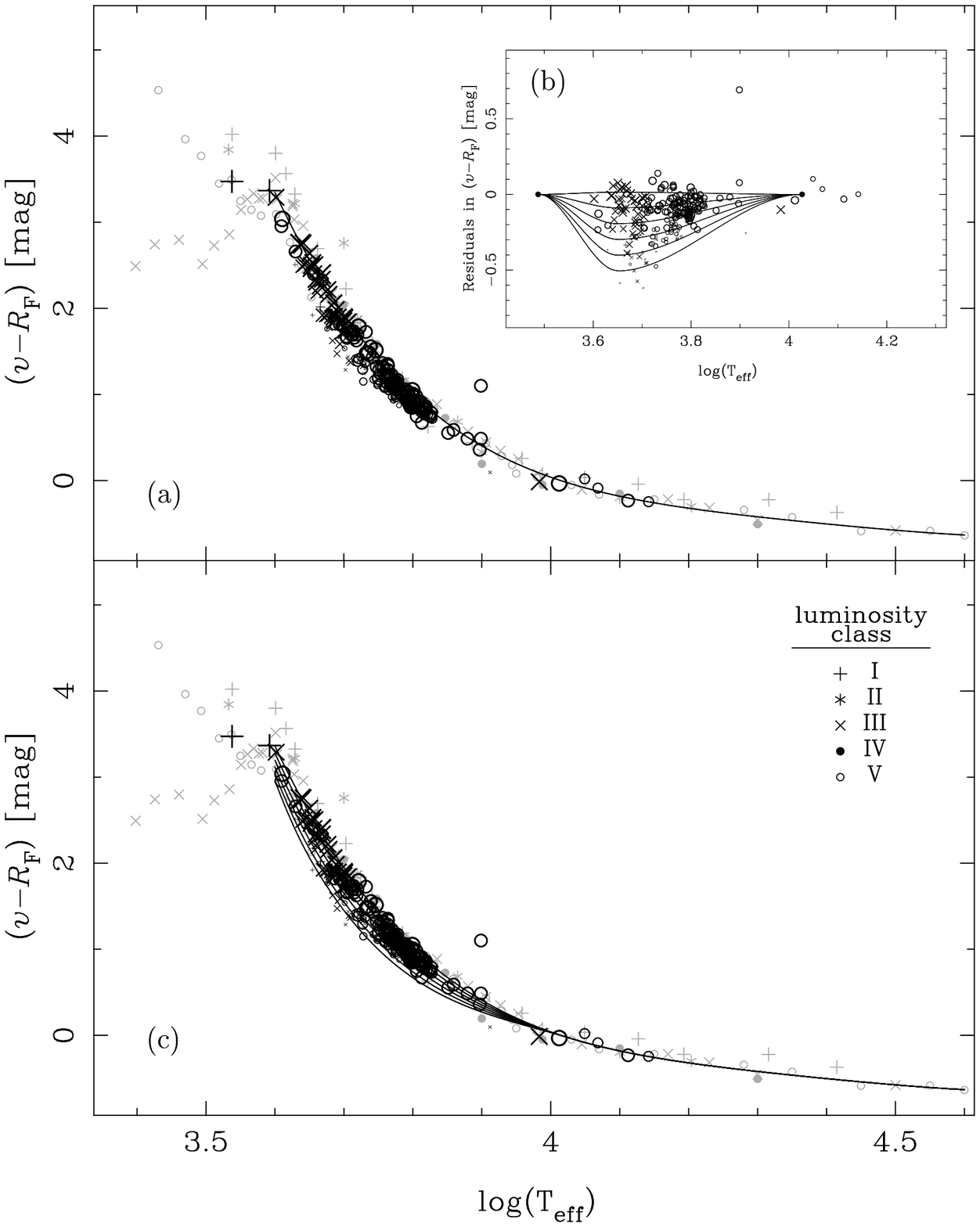}}
\caption{Variation of the $(v-R_{\rm F})$ colour as a function of effective
temperature. Different symbols correspond to distint luminosity classes, as
explained in the key, whereas symbol sizes are indicative of metallicity
(larger symbols for higher [Fe/H]). In panel~(a) and~(c) we have overplotted,
with light-grey symbols, the stellar sample of Pickles (1998). The
inclusion of the latter has allowed us to obtain an initial fit only dependent
on $T_{\rm eff}$ up to the highest temperature side. The inset in
panel~(b) corresponds to a zoom in the intermediate temperature interval (after
subtracting the initial fit), where the dependence on metallicity is more 
important. A new set of polynomials where fitted in this interval to
reproduce the behavior in both $T_{\rm eff}$ and [Fe/H] (where metallicities
are [Fe/H]=$-2.0$, $-1.5$,$-1.0$, $-0.5$, 0.0, and $+0.5$ from bottom to top). 
The combination of
these polinomyals and the initial fit leads to the final fitting functions
displayed in panel~(c). Note that these functions are not extrapolated below
$\log (T_{\rm eff}) < 3.6$.
\label{figure_colour_temperature}}
\end{figure}

Using all the possible pairs that can be built from any combination of the
previous 12 filters, we measured all these colours in the reddening free
stellar subsample. As it is expected, there is a clear variation of any of
these colours with effective temperature, with additional variations due to
metallicity for stars of intermediate temperature, and also to surface gravity.
An illustration of this behavior for the $(v-R_{\rm F})$ colour is shown in
Fig.~\ref{figure_colour_temperature}. Since for $\log(T_{\rm eff}) > 3.6$ and
$\log(g) \ge 3.5$ the effect of gravity is much less important than those of
temperature and metallicity, we have derived empirical fitting functions for
the colour variation as a function of only $T_{\rm eff}$ and [Fe/H] for these
temperature and gravity intervals. The fitting functions have been obtained
with three different sets of polynomials, forced to have common function values
and first derivatives at the joint points. The first and last polynomial sets
are only cubic functions on $T_{\rm eff}$, and the middle set is also cubic on
$T_{\rm eff}$ and linear on [Fe/H] (we have checked that no higher order in
metallicity is required to reduce the residual variance of the fits). In all
the cases, the two joint points of the three polynomial sets were also
considered as free parameters and were determined via the minimization
procedure of the fit. An example of these fitting functions are also plotted in
Fig.~\ref{figure_colour_temperature} for the $(v-R_{\rm F})$ colour.

\begin{figure}
\resizebox{1.0\hsize}{!}{\includegraphics[angle=-90]{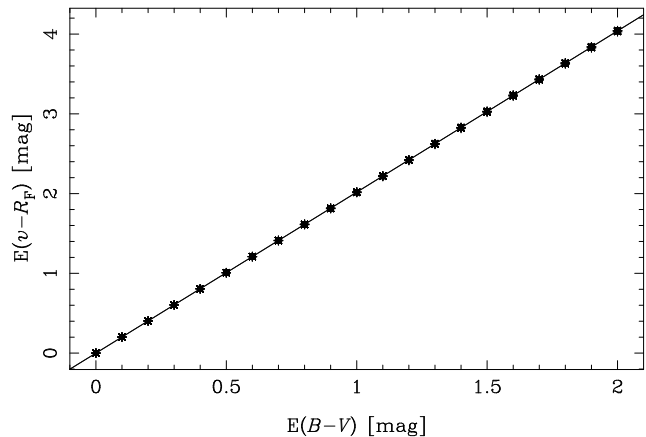}}
\caption{Comparison between the colour excess measured in the \mbox{($v-R_{\rm
F}$)} colour as a function of ficticious $E(B-V)$ artificially introduced in
the spectra of the reddening free stellar subsample. Although we are using the
same symbols than in Fig.~\ref{figure_colour_temperature}, it is clear that the
scatter introduced by distinct atmospheric stellar parameters is almost
negligible. The relation is nicely fitted by a second order polynomial.
\label{figure_ecolour_ebv}}
\end{figure}

\begin{figure}
\resizebox{1.0\hsize}{!}{\includegraphics[angle=-90]{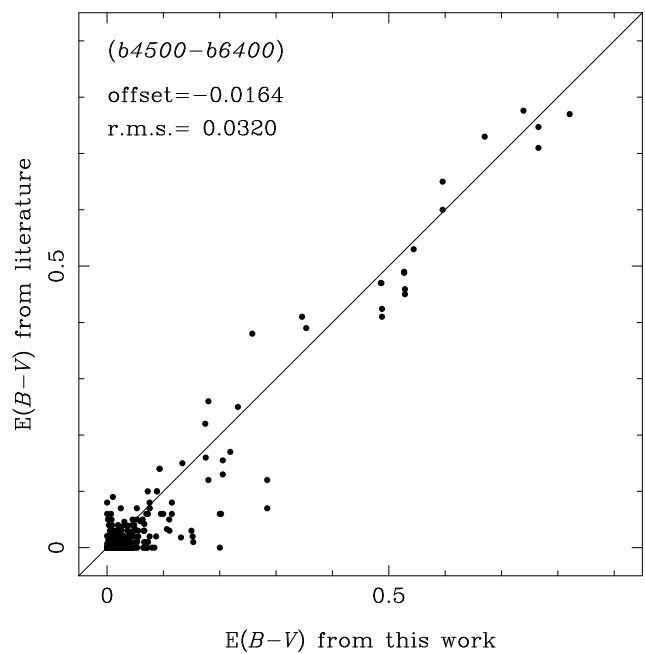}}
\caption{Comparison between $E(B-V)$ values from the literature with $E(B-V)$
estimations obtained from the excesses measured in the $(b4500-b6400)$ colour 
following the method described in this appendix.
\label{figure_ebv_ebv}}
\end{figure}

The derived fitting functions were then used to determine colour excesses of the
stars with known $E(B-V)$ from the literature. In order to perform this
computation, we previously obtained, empirically, the expected transformation
between the colour excess of any of the synthetic colours and the colour excess
in $(B-V)$. This was carried out by introducing ficticious reddenings in the
reddening free stellar subsample, parametrized as a function of $E(B-V)$, and
by measuring the corresponding excesses in the synthetic colour. In this step
we employed the Galactic extinction curve of Fitzpatrick (1999), with a ratio
of total to selective extinction at $V$ given by \mbox{$R_V=3.1$}. As it is
expected, and illustrated in Fig.~\ref{figure_ecolour_ebv}, for not very wide
filters there is an excellent correlation, almost independent on any stellar
atmospheric parameter, between reddenings estimated from different colours. By
measuring the differences between the fitting function predictions and the
actual synthetic colours of the stellar subsample with $E(B-V)$ from the
literature, and using the conversion between colour excesses, it was
straightforward to determine reddenings in $(B-V)$. The values obtained in
this way were compared with those from the literature.
The scatter of these comparisons were computed,
and the best 1:1 relation with the lowest scatter was obtained for the
$(b4500-b6400)$ synthetic colour, shown in Fig.~\ref{figure_ebv_ebv}

Once that we have determined the colour index that provided the best match with
the literature, we applied the method to transform
all the colour excesses in $(b4500-b6400)$ into $E(B-V)$ for the stars in the 
library whithout published values of this parameter (and in the ranges of
effective temperature and gravity above described).

  \end{document}